\documentclass[preprint]{aastex}

\usepackage{apjfonts} 
\usepackage{epsfig}

\newcommand\etal{et~al.} 

\slugcomment{Accepted for publication in \apjs}

\shorttitle{Yonsei Evolutionary Population Synthesis (YEPS) I.}
\shortauthors{Chung et al.}

\begin{document}

\title{YONSEI EVOLUTIONARY POPULATION SYNTHESIS (YEPS) MODEL. I.\\
SPECTROSCOPIC EVOLUTION OF SIMPLE STELLAR POPULATIONS}

\author{Chul Chung\altaffilmark{1, 2}, Suk-Jin Yoon\altaffilmark{1, 2}, Sang-Yoon Lee\altaffilmark{2}, \& Young-Wook Lee\altaffilmark{2}} 

\altaffiltext{1}{Equal first authors}
\altaffiltext{2}{Department of Astronomy \& Center for Galaxy Evolution Research, Yonsei University, Seoul 120-749, Republic of Korea; ywlee2@yonsei.ac.kr}

\begin{abstract}
We present a series of papers on the year-2012 version of Yonsei Evolutionary Population Synthesis (YEPS) model which is constructed on over 20 years of heritage.
This first paper delineates the {\it spectroscopic} aspect of integrated light from stellar populations older than 1~Gyr. 
The standard YEPS is based on the most up-to-date Yonsei-Yale stellar evolutionary tracks and BaSel 3.1 flux libraries, and provides absorption line indices of the Lick/IDS system and high-order Balmer lines for simple stellar populations as functions of stellar parameters, such as metallicity, age and $\alpha$-element mixture.
Special care has been taken to incorporate systematic contribution from horizontal-branch stars which alters the temperature-sensitive Balmer lines significantly, resulting in up to 5 Gyr difference in age estimation of old, metal-poor stellar populations. 
We also find that the horizontal branches exert an appreciable effect not only on the Balmer lines but also on the {\it metallicity-sensitive} lines including the magnesium index. 
This is critical to explain the intriguing bimodality found in index distributions of globular clusters in massive galaxies and to derive spectroscopic metallicities accurately from various indices.  
A full set of the spectroscopic and photometric YEPS model data of the entire parameter space is currently downloadable at http://web.yonsei.ac.kr/cosmic/data/YEPS.htm.
\end{abstract}


\section{INTRODUCTION}

The Evolutionary Population Synthesis (EPS) technique is a key tool for interpretation of integrated light from remote stellar systems. 
Based on stellar evolution theories, the EPS models place constraints on ages, chemical abundances and star formation histories of star clusters and galaxies 
\citep[e.g.,][]{Tins78, Bruz83, Arim87, Guid87, Buzz89, Bruz93, Bres94, Frit94, Wort94, Leth95, Park97,1997ApJ...486..201Y, Fioc97, Mara98, Vazd99a, Schu02, Thom03, Bruz03, Lee05b, Lee05, Schi07, 2009MNRAS.392..691C, 2010ApJ...710..421L, 2010MNRAS.404.1639V, 2010ApJ...712..833C, 2011MNRAS.412.2445P, 2011MNRAS.418.2785M, 2012MNRAS.tmp.2944P}. 
Combined with recent development in high precision observations, the EPS models are becoming more important for the analyses of various stellar populations in galaxies.

In a series of papers, we intend to present the Yonsei Evolutionary Population Synthesis (YEPS) model for the spectroscopic and photometric evolution of simple stellar populations (SSPs). 
This paper, as the first paper of the series, describes the {\it spectroscopic} aspect of our YEPS model. 
The model is constructed by the YEPS Fortran code package, which has been improved and exploited for the past 20 years 
by many studies related to ($a$) synthetic color-magnitude diagrams for individual stars \cite[e.g.,][]{LDZ90, LDZ94, Lee05b,Rey01,Yoon02,2006A&A...459..499K,Yoon08,Han09} and ($b$) synthetic integrated spectra for colors and absorption indices of simple and composite stellar populations \cite[e.g.,][]{Park97,Lee00,Lee05a,Rey05,Rey07,Rey09,Kavi05,Kavi07a,Kavi07b,Kavi07c,Ree07,Yoon06,2009gcgg.book..367Y,2009gcgg.book..381Y,2008MNRAS.389.1150S,Mieske08,Choi09,2011ApJ...740L..45C, 2011ApJ...743..149Y, 2011ApJ...743..150Y, 2012MNRAS.tmp.2946C}.  
The forthcoming Paper II (Yoon et al. 2012, in preparation) will present the photometric evolution of stellar populations.  The latter papers in the series will discuss the effect of the different choice of model ingredients and input parameters on the model, as well as the application of the YEPS for the early-type galaxies that have composite stellar populations.

The standard YEPS model has been constructed based on the Yonsei-Yale (Y$^2$) stellar evolution models (Kim \etal\ 2002; S. Lee \etal\ 2012, in preparation) and the BaSeL flux library (Westera \etal\ 2002).
The absorption-line index model employs the Lick/IDS system \citep{Burs84, 1985ApJS...57..711F, Wort94, Wort97, Schi07}, which defines 25 absorption lines produced by various elements at the surface of stellar atmosphere. 
The Lick/IDS system uses the spectra of nearly 460 stars to cover a wide parameter space of temperature, gravity, and metallicity \citep{Buzz92, Buzz94, Wort94, Wort97, Schi07}. 
The system, however, does not consider the grid of $\alpha$-elements enhancement, and thus the enhancement should be treated theoretically. 
We applied the $\alpha$-element correction terms by \citet{Korn05} to our model, following the schemes used by \citet{Trag00}, \citet{Thom03}, and \citet{Schi07}.

The YEPS model has been built with particular interest 
in dealing with the core helium burning horizontal-branch (HB) stars.
Since the presence of hot stars in globular clusters (GCs) and galaxies 
changes the overall shape and absorption feature of their spectral energy distributions (SEDs), 
especially at short wavelengths, 
the impact of hot HB stars ($\gtrsim$\,8000\,K) has been a topic of great interest in the EPS community over the past 20 years 
\citep{LDZ90, LDZ94,1994ApJS...95..107W, Lee00, Thom03, Lee05b, Schi07, Yoon06, Yoon08, 2011ApJ...743..149Y, 2011ApJ...743..150Y,yoon12}. 
\citet{Lee00} first demonstrated that H$\beta$ absorption index---the most popular age indicator---is significantly enhanced 
by the presence of blue HB stars. 
More recently, \citet{Yoon06, 2011ApJ...743..149Y, 2011ApJ...743..150Y} show 
that the systematic metallicity-dependent variation in HB temperature 
leads to the nonlinear relationship between metallicity and broadband optical colors.
Despite the fact that hot, blue HB stars exert a strong effect on properties of integrated light from GCs and galaxies, 
most EPS models to date took into account the details of HBs in a fairly limited manner.
The YEPS, by contrast, elaborates the HB effect not as a merely contamination factor 
but as a crucial part of the EPS model for various spectroscopic and photometric observables.

The paper is organized as follows. Section 2 describes the constructing procedure of the YEPS model.
Section 3 presents the results of our stellar population simulations and the comparison of our model with observations. 
Section 4 discusses the implications, and finally Section 5 summarizes our results.
A full set of the spectroscopic and photometric YEPS model data of the entire parameter space is available at http://web.yonsei.ac.kr/cosmic/data/YEPS.htm.

\vspace{0.2cm}
\section{CONSTRUCTION OF THE YEPS MODEL}

The YEPS model provides, for a given stellar system, ($a$) the synthetic color-magnitude diagrams (Section 2.1), ($b$) the synthetic spectral energy distributions (Section 2.2), ($c$) the integrated absorption line indices (Section 2.3), ($d$) the integrated magnitudes and broadband colors, and ($e$) the integrated surface bright fluctuations. 
Table~\ref{tab.1.1} summarizes the ingredients and input parameters of the YEPS model.

\subsection{Synthetic Color-Magnitude Diagrams}

The standard YEPS model is constructed based on the most up-to-date Yonsei-Yale (Y$^2$) stellar evolutionary tracks. 
For the evolutionary phases from the main sequence (MS) to the tip of the red giant branch (RGB), we used Y$^2$-isochrones (\citealt{Kim02}; Lee et al. 2012, in preparation), covering the metallicity grids from $Z$ = 0.00001 to 0.08 with three different values for the $\alpha$-elements enhancement ([$\alpha$/{\rm Fe}] = 0.0, 0.3, and 0.6).
The mixture pattern of $\alpha$-elements enhancement in Y$^2$-isochrones follows that of \citet{2000ApJ...532..430V}. 
The Y$^2$ stellar evolutionary libraries adopt the galactic helium enrichment parameter of $\Delta Y/\Delta Z = 2.0$ with the primordial helium abundance of $Y = 0.23$. 
To examine the effect of the different choice of the evolutionary tracks, we comparatively used the BaSTI stellar evolutionary tracks \citep{Piet04} with metallicities from $Z = 0.0001$ to 0.04 for the two $\alpha$-elements enhancement cases ([$\alpha$/{\rm Fe}] = 0.0 and 0.4).
{The Y$^2$ stellar libraries include helium diffusion and BaSTI stellar libraries include the atomic diffusion of both helium and metals.}
As will be demonstrated below, the major features of our model do not depend on the specific choice of stellar libraries.

We adopt the Salpeter initial mass function (IMF) for our standard set of simulations to assign the number of stars along given isochrones. 
\citet{Wort94} presented the generalized Salpeter IMF of the form
\begin{equation}
\frac{dN}{dM} = \frac{M_{tot} (1-x)}{M_u ^{1-x} - M_l ^{1-x}}M^{-(1+x)} ,
\label{eq.1.1}
\end{equation}
where $dN$ is the number of stars within the fixed mass bin $dM$, and $M_l$ and $M_u$ are the lower and upper mass cuts, respectively. 
From this, we calculated the IMF of an SSP that consists of an single-metallicity and single-age population. 
{We have applied 10$^6$ stars within whole mass range of IMF (from 0.2 to 5.0 $M_{\odot}$).}
We adopted the standard Salpeter index ($x = 1.35$) over the whole mass range.
The choice of $x$ exerts a fairly small effect on the overall shape of UV-to-IR SEDs and hence on broad-band colors and absorption indices \citep{Park97}.
This is more so for old stellar populations for which massive stars are already evolved off the MS
because the index $x$ controls the fractional contribution from the massive stars. 
However, it is noteworthy that $x$ leads to significant variations in total absolute magnitude of the model SSPs \citep{1972ApJ...178..319T} 
because the absolute flux level of SEDs is a function of the total stellar mass.

For the synthetic HB modeling, we used Y$^2$-HB tracks (Lee et al. 2012, in preparation) that are fully consistent with Y$^2$-isochrones in terms of the input physics and assumed parameters. 
{The Y$^2$ HB tracks cover a wide range of HB total mass from 0.4438 $M_{\odot}$ for Z=0.06 and 0.5037 $M_{\odot}$ for Z=0.0001 to 1.5 $M_{\odot}$ for all metallicities to incorporate the wide variation of HB morphology.}
In order to simulate the mass dispersion of HB stars, we have used the Gaussian HB mass distribution of the form 
\begin{equation}
P(M) \propto exp \left( \frac{-(M- \left< M_{HB} \right>)^2}{2\sigma_M ^2} \right),
\label{eq.1.2}
\end{equation}
where $P(M)$ is the probability density function of the HB mass, and $\left< M_{\rm HB} \right>$ is the mean mass of the HB at a given metallicity and age. 
The standard model assumed the value of $\sigma_M$ to be $0.015M_{\odot}$ \citep{LDZ90, LDZ94}.
{
On average, the number of HB stars at given metallicity and age is 350 in a single simulation. In addition, in order to avoid small number statistics in HB modeling, we have repeated the simulation 10 times to get the averaged continuum flux at given metallicity and age.
}

\begin{figure*}
\includegraphics[angle=-90,scale=0.8]{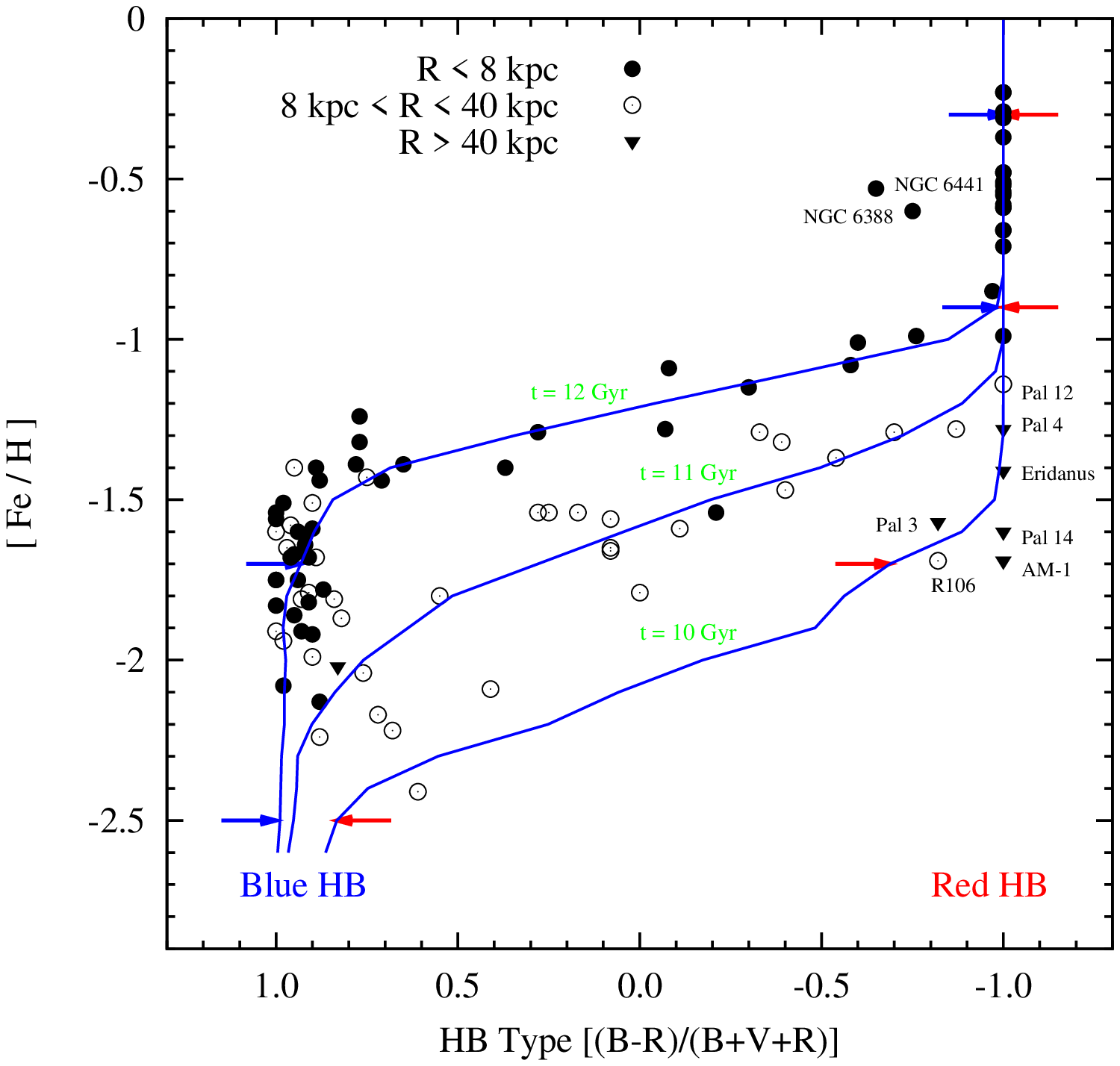}
\caption[]{The variation of HB morphology as functions of metallicity and age. Filled and open circles and triangles are GCs in three radial zones of the Milky Way \citep{LDZ94, 2007ApJ...661L..49L}. 
Solid blue lines are theoretical isochrones from 10, 11, and 12 Gyr YEPS model.
Blue and red arrows indicate the selected metallicities ([Fe/H] = $-2.5$, $-1.7$, $-0.9$, and $-0.3$) for synthetic color-magnitude diagrams (CMDs) in Figure~\ref{1.2}.}
\label{1.1}
\end{figure*}

\begin{figure*}
\includegraphics[angle=-90,scale=0.85]{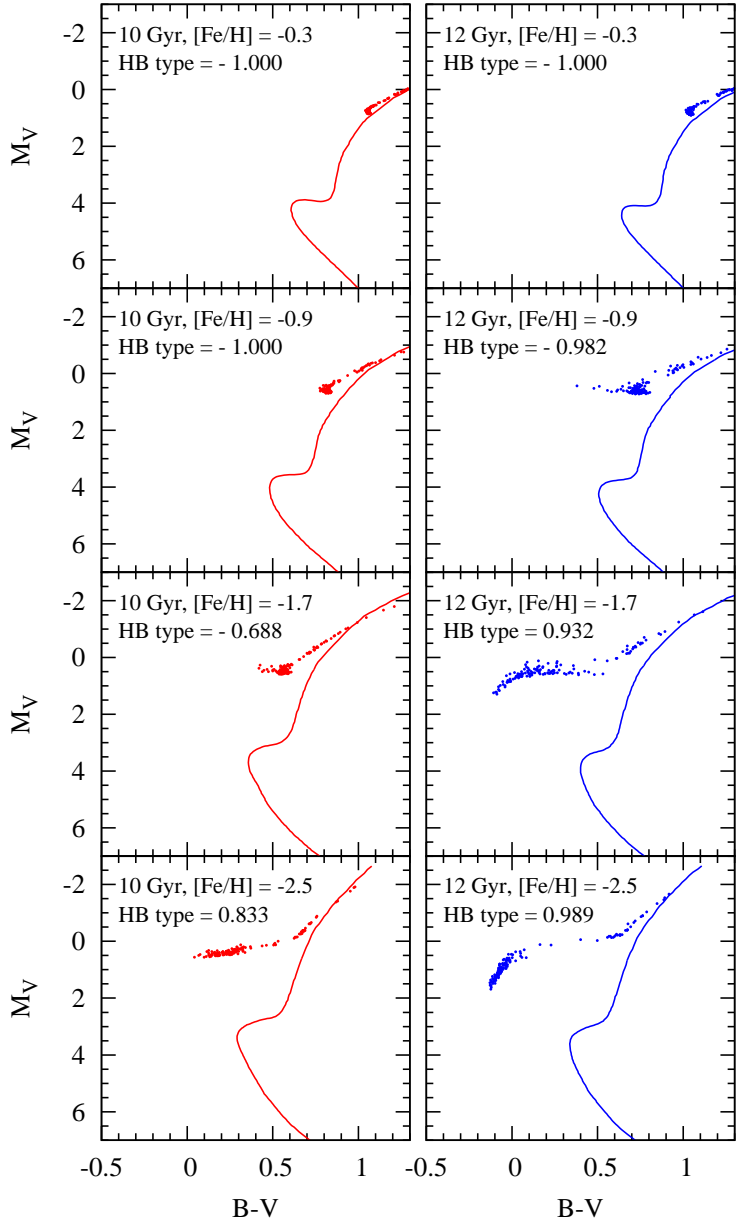}
\caption[]{The effect of metallicity and age on the morphology of HB stars in the CMD. 
Metallicity of each CMD increases from the bottom to the top panels. CMDs on the left and right sides are 10 and 12 Gyr 
models, respectively. Solid lines are isochrones and dots are corresponding HB stars at given ages and metallicities.}
\label{1.2}
\end{figure*}

Figure~\ref{1.1} shows how the HB morphology of YEPS model\footnote{Recent observations and modeling indicate that the abundance anomaly, especially in He and CNONa, is present in the MWGCs with multiple stellar populations. Although the variation in He and CNONa abundance among GCs in the MW is large, the average variation between GC systems in different galaxies, as a whole, is not expected to vary greatly. Additionally, only 30~\% of the MWGCs are significantly affected by the enhancement in He \citep{2007ApJ...661L..49L}, and this suggests that HB morphologies in the majority of the MWGCs are mostly controlled by total metallicity and age. However, if the average He enhancement in the MWGCs is not archetypical, and it varies significantly from one galaxy to another, our models presented here would need further revisions to reflect this.} is calibrated to the observations.
The HB type is defined as $(B-R)/(B+V+R)$, where $B$, $R$ and $V$ are the numbers of blue and red HB stars and RR Lyrae variable stars, respectively \citep{LDZ94}.
Filled circles represent the oldest inner-halo ($R_{GC}$\,$<$\,$8$ kpc) the Milky Way globular clusters (MWGCs), and open circles ($8$\,$<$\,$R_{GC}$\,$<$ $40$ kpc) and triangles ($R_{GC}$\,$>$\,$40$ kpc) represent the outer-halo MWGCs. 
Solid lines from top to bottom are the HB type variation of the YEPS model with varying ages. 
The free parameter, the Reimers mass-loss efficiency parameter $\eta$, is used to calibrate our model HB types to the observations.
We adopted \citet{Reim77}'s empirical formula for the mass loss along the RGB \citep{Rood73, LDZ90}. 
The formula takes the form of $\frac{dM}{dt} \propto \eta \frac{L}{gR}$, where $L$, $g$, and $R$ are luminosity, gravity, and radius of stars, respectively.
The comparison of models and observations suggest a $\eta$ of 0.63 under the assumption that the mean age of inner-halo GCs is 12 Gyr \citep{Grat97, Reid97, 1998ApJ...494...96C, 2009ApJ...694.1498M, 2010ApJ...708..698D}. 
Figure~\ref{1.2} shows the color-magnitude diagrams (CMDs) for selected model SSPs (red and blue arrows in Figure~\ref{1.1}). 
In general, the HB type becomes redder with increasing metallicity and decreasing age \citep[e.g.,][]{LDZ94}. 

\subsection{Synthetic Spectral Energy Distributions}

Spectral energy distributions (SEDs) of SSPs are generated based on the synthetic CMDs (Section 2.1).
The CMDs give the stellar parameters of individual stars in given SSPs, including effective temperature ($T$), surface gravity ($g$), global metallicity ($[{\rm Z}/{\rm H}]$), and $\alpha$-element enhancement ([$\alpha$/{\rm Fe}]). 
To derive theoretical spectral fluxes (in units of [erg/s/cm{$^2$}/{\small \AA}]), we use the spectral library of BaSel 3.1 \citep{West02}. 
BaSel 3.1 is based on the expertise of \citet{Kuru92} and BaSel 2.2 \citep{Leje98}, and provides extensive and homogeneous grids of theoretical flux distributions calibrated to {the colors of the MWGCs at all levels of metallicity.} 
The library covers effective temperatures from 2,000\,K to 50,000\,K, gravities in a solar unit from $\log g$ of $-1.02$ to 5.50, and metallicities $[{\rm Z}/{\rm H}]$ from $-$2.0 to 0.5. 
Note that the BaSel 3.1 library assumes scaled-solar $\alpha$-elements. 
We thus choose to use the total metallicity $[{\rm Z}/{\rm H}]$ for the construction of SEDs of SSPs, rather than the iron abundance [Fe/H]\footnote{In some photometric broadband colors (e.g., $U-B$), broadband colors for $\alpha$-enhanced mixture are better reproduced by scaled-solar spectra with the same $[{\rm Fe/H}]$ of the $\alpha$-enhanced mixture, not $[{\rm Z/H}]$ \citep{2004ApJ...616..498C}.}. 
The equation $[{\rm Z}/{\rm H}]=[{\rm Fe}/{\rm H}]+A\,[\alpha/{\rm Fe}]$ relates [Z/H] to [Fe/H].
In our model for $[\alpha/{\rm Fe}]$ = 0.3, the factor $A$ which depends on the $\alpha$-element mixture of the model equals to 0.723. 

We calculate the expected flux $F_{\lambda}$ at a distance $d$ using the form 
\begin{equation}
F_{\lambda} = 4\pi \times \frac{L}{\sigma T_{\rm eff}^4} \times H_{\lambda} \times \frac{1}{4\pi d^2} ,
\label{eq.1.3}
\end{equation}
where $H_{\lambda}$, $L$, $T_{\rm eff}$ and $\sigma$ are the flux intensity, luminosity, effective temperature of a star, and the Stefan-Boltzmann constant, respectively. 
From these fluxes of individual stars, integrated fluxes of all stages of stellar populations {in the synthetic CMDs}---MS to RGB, HB and post-asymptotic giant branch (PAGB)---are calculated using the following summation form:

\begin{equation}
F_{\lambda}^{total} = F_{\lambda}^{MS} + F_{\lambda}^{RGB} + F_{\lambda}^{HB} + F_{\lambda}^{PAGB}.
\label{eq.1.4}
\end{equation}
{Total mass of SSPs at given age and metallicity is normalized to $10^6$ $M_{\odot}$.}

\vspace{0.1cm}
\subsection{Absorption-line Strength Indices}

The absorption-line indices of the YEPS model are calculated using the polynomial {\it fitting functions}. 
The fitting functions are derived from spectra (in the 4,000\,--\,6,000\,{\small \AA} region) of Galactic stars 
and yield the line strengths as functions of stellar atmospheric parameters---metallicity, temperature, and gravity \citep{Rose85, Jone95, Vazd99a, Fabe73, Rose84, Diaz89, Wort94, Buzz92, Buzz94, Wort97, Schi07}. 
The empirical polynomial fitting functions, combined with continuum levels of model SEDs, 
generate the absorption-line indices of SSPs.

\begin{figure*}
\includegraphics[angle=0,scale=0.8]{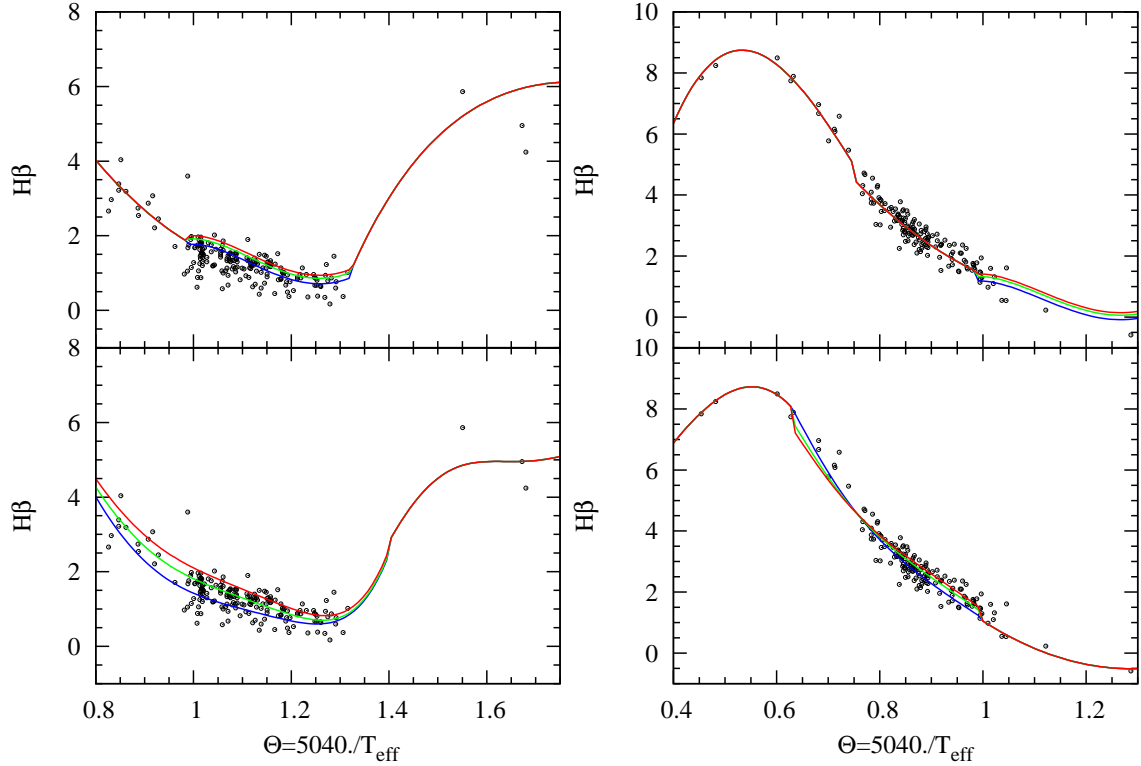}
\caption[]{ H$\beta$ fitting functions with respect to the gravity and temperature ($\Theta$ = 5040/T$_{\rm eff}$). 
Black dots are H$\beta$ measurements of 460 stellar spectra used for the construction of W94's fitting functions.
Left panels are fitting 
functions for giant stars (with $\log g \leq 3.6$) and right panels are fitting functions for dwarf stars (with $\log g \geq 3.6$). Upper and bottom 
panels are fitting functions of W94 and S07, respectively. Red, green, and blue solid lines indicate metallicities of 
[Fe/H] = $-1.0$, $-0.3$, and 0.1, respectively.}
\label{1.3}
\end{figure*}

For the standard YEPS absorption index model, we use \citet{Wort94} and \citet{Wort97}'s (hereafter W94 and W97) polynomial fitting functions for 25 absorption indices of the Lick/IDS system. 
As a comparison model, we adopt \citet{Schi07}'s (hereafter S07) fitting functions in the blue wavelength based on \citet{Jone99} stellar library. 
Figure~\ref{1.3} shows an example of the fitting functions (H$\beta$ line) as given in W94 and S07. 
The two sets of fitting functions agree well with each other. 
We note that the H$\beta$ fitting function of S07 exhibits greater metallicity-sensitivity for giant stars than that of W94
by virtue of a more recent spectral library by \citet{Jone99}.

From these fitting functions, we derive the equivalent width of a single star at a given temperature, gravity, and metallicity. 
The empirical fitting function provides an absorption index ($I$), which is transformed to the equivalent width (EW) using the following equations;
\begin{equation}
 EW({\rm \small \AA})=f_C \times \left( \frac{1-I}{\Delta\lambda} \right),
\label{eq.1.5}
\end{equation}
\begin{equation}
 EW(mag)=f_C \times 10 ^{-0.4 I },
\label{eq.1.6}
\end{equation}
where $\Delta \lambda$ and $f_C$ are the index bandpass and flux continuum of an SED for each index. 
{The continuum fluxes ($f_C$) of each index were taken from the definition of Lick standard system described in W94.}
The unit of $mag$ is used for CN$_{1}$, CN$_{2}$, Mg$_{1}$, Mg$_{2}$, TiO$_{1}$, and TiO$_{2}$ lines.

After determining the EW and $f_C$ of individual model stars, we finally calculate the integrated indices by summing the continua and EWs of all stellar populations of SSPs using the following formulae, 
\begin{equation}
\begin{array}{l}
Integrated\\
i^{th}~ Index\\
\end{array}
={\left\{
\begin{array}{l}
  \Delta \lambda^i \times \left( 1- \frac{\sum_{j}^{}\left[f_{C,j}\times \left( 1 - \frac{I_j}{\Delta \lambda^i}\right)\right]}
{\sum_j^{} f_{C,j}}\right)
\\
-2.5 log \left(\frac{\sum_{j}^{}\left[f_{C,j}\times 10^{\left( -0.4 I_j \right)}\right]}{\sum_{j}^{} f_{C,j}}\right) 

\end{array} 
\right\}}
,
\label{eq.1.7}
\end{equation}
where $f_{C,j}$, $I_j$, and $\Delta \lambda^i$ are the $j^{th}$ model star's continuum, absorption index, and the width of the $i^{th}$ index bandpass, respectively.

\subsection{Treatment for the Enhancement of $\alpha$-elements }

The absorption indices for the case of enhanced $\alpha$-elements are modeled as follows. 
We use Y$^2$ stellar evolutionary tracks with enhanced $\alpha$-elements \citep{Kim02}. 
For the stellar atmosphere model, we adopt the $\alpha$-elements mixture ratios that is identical to that of stellar tracks for the sake of consistency.
Table~\ref{tab.1.2} shows the $\alpha$-elements mixture of the Y$^2$ stellar evolutionary model and BaSTI, 
as compared with the scaled-solar abundance ratio of metals taken from \citet{1993oee..conf...15G}. 
Our YEPS models for enhanced $\alpha$-elements follow the two $\alpha$-element mixtures of Y$^2$ and BaSTI.

We then applied the response functions of $\alpha$-elements by \citet{Korn05} (hereafter K05) to apply the $\alpha$-elements mixture to YEPS. 
Compared to the previous work by \citet{Trip95}, K05 have a more extended metallicity space ranging from $[{\rm Fe}/{\rm H}]$ = $-$2.25 to $+$0.67, and provide the response functions for 25 Lick absorption indices for three evolutionary phases (dwarfs, turn-offs, and giants). 
The response functions that K05 provide are the first partial derivative $\partial I / \partial[X_i ]$ of the Lick index $I_0$ when an abundance increment of two times logarithmic $i^{th}$ $\alpha$-element (C, N, O, Mg, Fe, Ca, Na, Si, Cr and Ti) is assumed. 
As described in TMB03, it is appropriate to expect $I \propto \exp ([X_i ])$ for the optimal approximation. 
Hence, the Taylor expansion for $\ln I$ instead of $I$ is an adequate approach for the variation of the Lick absorption indices due to $\alpha$-element abundance changes. 
Neglecting the higher-order derivatives and following the notation $R_{0.3} (i)$ of \citet{Trag00}, we can express the Taylor expansion in the following forms, 
\begin{equation}
\begin{array}{l}
\ln I_{new}\\
~
\end{array}
\begin{array}{l}
 = \ln I + \sum_{i=1}^{n} \frac{\partial{\ln I}}{\partial{[X_{i}]}}\\
= \ln I + \sum_{i=1}^{n} R_{0.3} (i) \frac{\Delta [X_i ]}{0.3} ,
\end{array}
\label{eq.1.8}
\end{equation}
where $R_{0.3}(i)={1}/{I_0 }\times{\partial I}/{\partial [X_i ]}\times0.3$ is the K05 index response for increased $\alpha$-element $i$ by 0.3 dex, and $I$ and $I_0$ are the absorption index before applying the K05 response function and the model absorption index of the K05 parameter space, respectively. 
The exponential scale of Eqn.~\ref{eq.1.8} yields

\begin{equation}
I_{new} = I \prod_{i=1}^{n} {\exp \left(R_{0.3} (i) \right)}^{\left(\frac{\Delta{[X_{i}]}}{0.3}\right)}.
\label{eq.1.9}
\end{equation}

Fitting functions of W94 and S07 give negative values for absorption indices when stellar populations are young, metal-poor (e.g., CN, Ca, and Fe lines) or old, metal-rich (e.g., H$\beta$, H$\gamma$, and H$\delta$). 
Since our calculation of the $\alpha$-element fractional change is based on the logarithmic Taylor series of $\ln I$, we must avoid negative values of the absorption indices. 
Table~\ref{tab.1.3} lists the negative minimum values of YEPS absorption index models when we adopt W94 fitting functions together with Y$^2$ and BaSTI libraries.
The simplest way to correct negative values in the fractional index change is to shift the negative indices into the zero or positive value and then compute the fractional change. 
We used the correction term $\delta$ as listed in Table~\ref{tab.1.3}, and applied the following equation to correct absorption indices with negative values,
\begin{equation}
I_{new}-\delta = (I-\delta) \prod_{i=1}^{n} {\exp\left(\frac{1}{I_{0}-\delta}\frac{\partial{I}}
{\partial{[X_{i}]}0.3}\right)}^{\left(\frac{\Delta{[X_{i}]}}{0.3}\right)}.
\label{eq.1.10}
\end{equation}
After the derivation of $I - \delta$, we scale back the index $I_{new} - \delta$.

\vspace{0.3cm}

\section{RESULT OF STELLAR POPULATION SIMULATIONS}

\subsection{The Effect of HB Stars on Simple Stellar Populations}

\begin{figure*}
\includegraphics[angle=-90,scale=0.79]{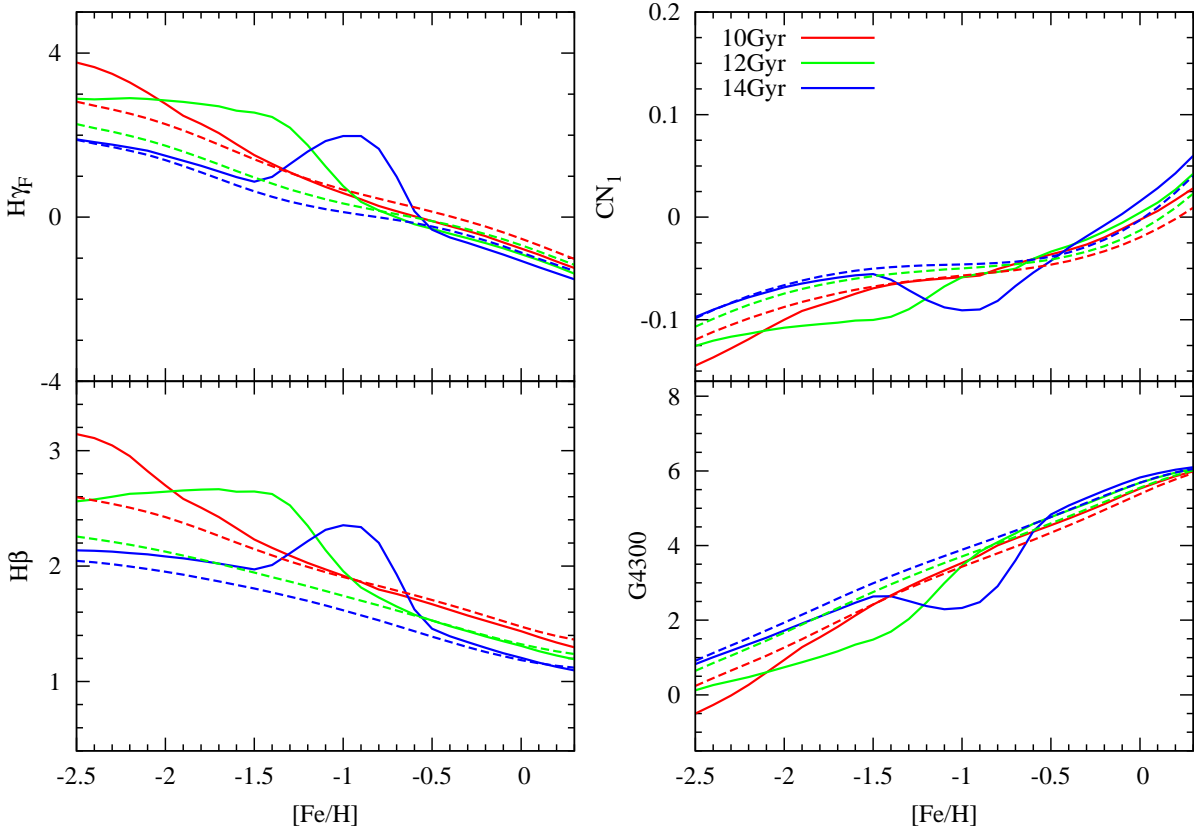}
\caption[]{The effect of HB stars on the absorption indices of CN$_{1}$, G4300, H$\gamma_{F}$, and H$\beta$. 
Red, green, and blue lines indicate, respectively, SSP models with ages of 10, 12, and 14 Gyr. 
Solid and dashed lines are YEPS models with and without HB stars, respectively. 
The models are for [$\alpha$/Fe] = 0.3.}
\label{1.4}
\end{figure*}

Figure~\ref{1.4} shows the most temperature-sensitive indices (Balmer lines, CN$_{1}$, and G4300) as a function of [Fe/H] and highlights the effect of HB stars on integrated absorption line strengths of SSPs.  In the YEPS model without HBs, H$\beta$ index decreases monotonically with increasing [Fe/H] at given age. 
This trend is due solely to the temperature variation of turn-off (TO) and RGB stars, which becomes cooler as the metallicity of stellar population increases.
By contrast, in the models with HB stars (solid lines), the H$\beta$ and H$\gamma_{F}$ strengths are significantly enhanced by the presence of blue HB stars at the metal-poor regime.
CN$_{1}$ and G4300 are also sensitive to hot HB stars (W94) in the sense that these lines get weakened by the presence of hot HB stars.
The models with even hotter HB stars in the most metal-poor GCs tend to approach the models without HBs, as the hotter HB stars are dim and too hot to have significant effect on these indices. 

The fact that blue HB stars can mimic young, hot TOs in the integrated fluxes of GCs has significant implication for determining ages and metallicities of stellar populations. 
For instance, when the H$\beta$ strength is used as an age indicator, a 7 Gyr model with no HB stars exhibits H$\beta$ strength identical to a 12 Gyr model with HB stars at [Fe/H] $\simeq$ $-1.6$.
This suggests that one could seriously underestimate age of stellar populations for old ($>$ 10 Gyr) populations. 
Also, HB stars affect other absorption indices known as metallicity indicators
(e.g., Mg\,$b$, Fe4383, and $\left< \rm Fe\right>$), although the amount of change is small compared to the Balmer indices (see Section 4 below).

\begin{figure*}
\includegraphics[angle=0,scale=0.85]{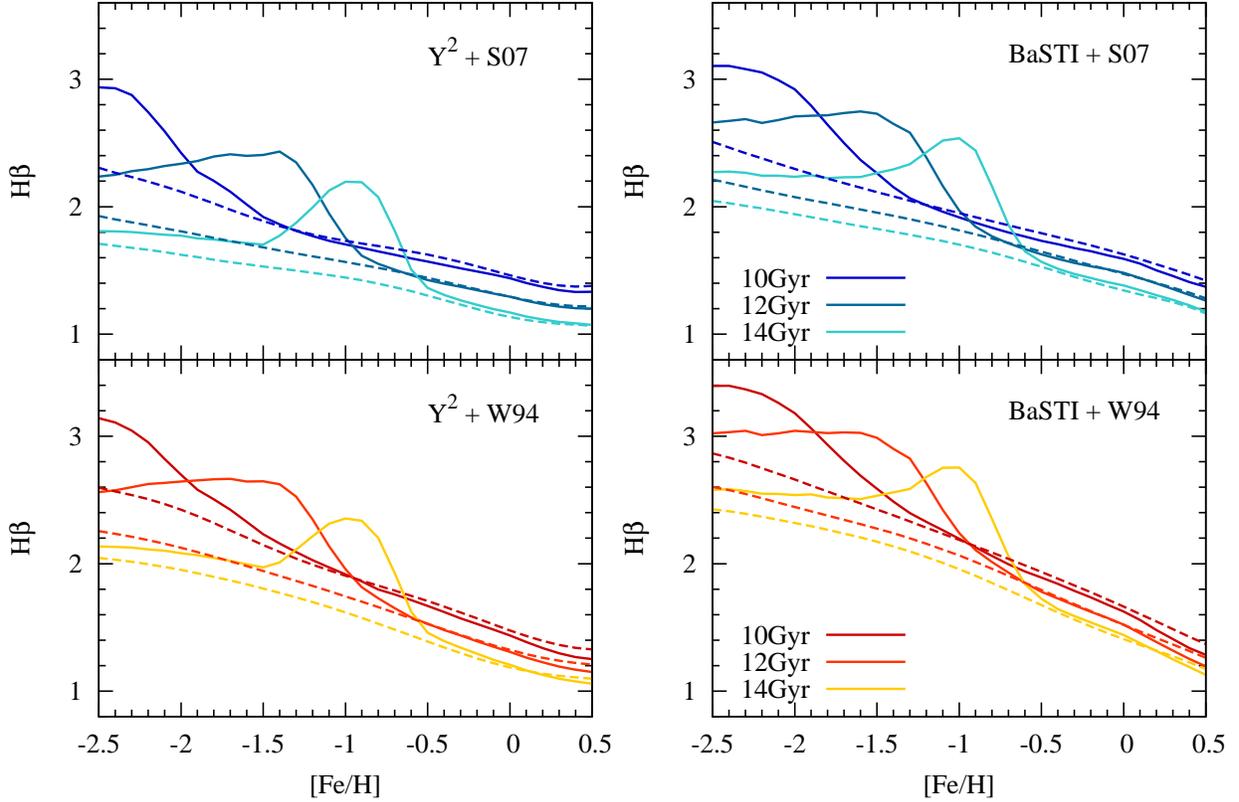}
\caption[]{The effect of various input parameters (stellar libraries and fitting functions) on the strength of H$\beta$. 
Three color codes indicate different ages of 10, 12, and 14 Gyr. 
Solid and dashed lines are YEPS models with and without HB stars. 
Models in upper panels are calculated based on fitting functions of S07 and models in bottom panels are constructed using fitting functions of W94.
Stellar evolutionary tracks used in the left and right panels are Y$^2$ \citep{Kim02} and BaSTI \citep{Piet04} stellar libraries, respectively.
}
\label{1.5}
\end{figure*}

Given the wide implication of the HB effect, it is highly important whether or not such an effect depends on the specific choice of stellar libraries for evolutionary tracks and the fitting functions.
In this study, we employ two stellar libraries (Y$^2$ and BaSTI) and two empirical fitting functions (W94+W97 and S07). 
The four combinations of the models are presented in Figure~\ref{1.5}. 
Solid and dashed lines represent the H$\beta$ strengths for the models with and without the HB prescription. 
Comparison shows that all models agree well for the H$\beta$ index.
For instance, all the 12 Gyr models with HB stars show the same amount of H$\beta$ enhancement of about 0.6\,{\small \AA} at [Fe/H] $\simeq$ $-1.6$ compared to the model without HB stars.
Our additional test using the high-resolution spectra of \citet{2005A&A...442.1127M} also confirms that, compared to the model without HBs, the model with HBs at the same condition shows enhanced H$\beta$ by about 0.6\,{\small \AA}.
We note that comparison between models without HB stars (dashed lines) shows that the Y$^2$ stellar library produces a larger gap between iso-age lines than the BaSTI library.
The reason for this is that the Y$^2$-isochrones have larger TO temperature gaps than the BaSTI isochrones.
In addition, the fitting functions of S07 yield slightly weaker H$\beta$ indices in the metal-poor regime than W94.

\subsection{The Effect of $\alpha$-elements on Simple Stellar Populations}

\begin{figure*}
\includegraphics[angle=0,scale=1.0]{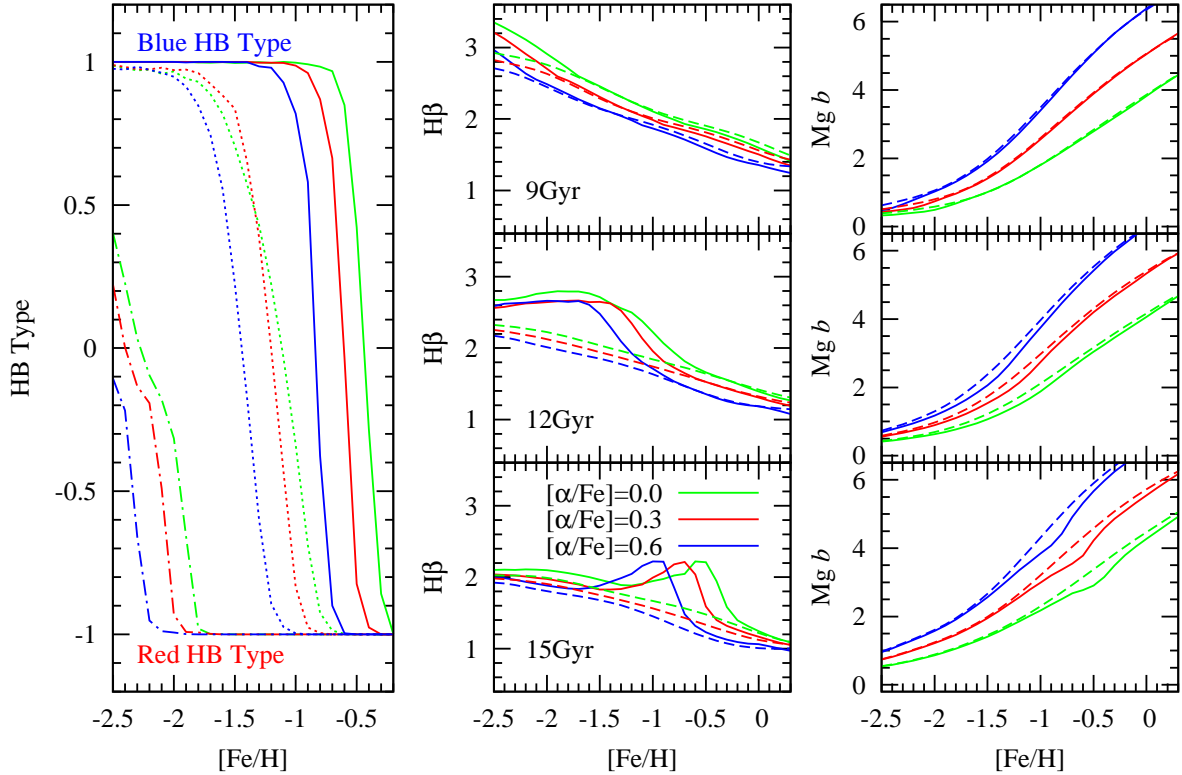}
\caption[]{The effect of $\alpha$-elements on the HB type, H$\beta$, and Mg\,$b$ indices.
Green, red, and blue colors are the YEPS models for [$\alpha$/Fe] = 0.0, 0.3, and 0.6, respectively. 
The left panel is the same plot as Figure~\ref{1.1} but for various $\alpha$-element enhancements. 
Solid, dotted, and dot dashed lines in the left panel are HB types for 3 different ages of 15, 12, and 9 Gyr, respectively. 
Middle panels are the H$\beta$ models with and without HB stars for different ages. 
Solid and dashed lines in the middle and right panels represent the YEPS models with and without HB stars. 
From top to bottom panels in the middle and right panels, the ages of the YEPS models are 9, 12, and 15~Gyr.
Right panels are the YEPS models for the Mg\,$b$ index with the same parameters as in the middle panels.}
\label{1.6}
\end{figure*}

Figure~\ref{1.6} demonstrates the effect of $\alpha$-elements on Lick indices using two indices (H$\beta$ and Mg\,$b$) that are sensitive to the effective temperature and $\alpha$-elements, respectively.
The middle column shows the effect of $\alpha$-elements on H$\beta$ as functions of [Fe/H] and age.
The strength of H$\beta$ index without HB stars (dashed lines) decreases with increasing $[\alpha/{\rm Fe}]$.
This is because temperature of TO stars decreases with increasing [$\alpha$/Fe].
Hence, the use of stellar libraries that incorporate enhanced $\alpha$-elements is crucial to predict accurately the strength of H$\beta$.
The total $Z$ increment due to the $\alpha$-elements enhancement also exert effect on the H$\beta$ index with HB stars.
The way the $\alpha$-elements affect the HB types is shown in the left column of Figure~\ref{1.6}. 
As [$\alpha$/Fe] increases, the models with enhanced [$\alpha$/Fe] (blue lines) show redder HB types compared to the scaled solar models (green lines) at given [Fe/H].
The strengths of H$\beta$ index in the middle column reflect the trends on the HB types shown in the left column. 
Interestingly, the 15 Gyr model (solid lines in the left column) has blue HB type at [Fe/H] = $-$2.5 $\sim$ $-$1.5, but the contribution of blue HBs to H$\beta$ is small compared to the 12 Gyr model at same metallicity.
The first reason for this is that the strength of H$\beta$ reaches its maximum at $T_{\rm eff} \simeq$ 9,500\,K. 
So, the contribution from hotter ($T_{\rm eff} >$ 9,500\,K) HBs in older and/or more metal-poor populations to the H$\beta$ absorption becomes smaller. 
The second reason is that those HBs with hot temperature are fainter. 
Hence, their contribution to luminosity-weighted absorption indices becomes relatively small.

It is important to note that Mg\,$b$, the well-known tracer of [Fe/H] and [$\alpha$/Fe],  is also affected by the variation of HB morphologies in their absorption strengths (see 12- and 15-Gyr models).
To verify the effect of HB stars on Mg\,$b$, we have calculated the absorption strength for the following two cases. 
The fitting function of S07 instead of W94 yields, for 12-Gyr SSP with [Fe/H] $\simeq$ $-1.0$, the Mg\,$b$ strength that is decreased by 0.28\,{\small \AA} due to the effect of HB stars.
The line strengths produced by high-resolution spectra of \citet{2005A&A...442.1127M} shows that Mg\,$b$ decreases by 0.36\,{\small \AA} under the same condition. 
The results imply that the effect of HB stars on Mg\,$b$, albeit being relatively small compared to H$\beta$, 
is not negligible and the use of Mg\,$b$ as a direct tracer of metallicity 
should be with more caution and require modification. 
It is clear that, even after the enhanced $\alpha$-elements are applied to SSPs, 
the HB effect on the absorption indices still dominates the metal-poor regime irrespective of the [$\alpha$/Fe] values. 

\subsection{Comparison with Other Models}

\begin{figure*}
\includegraphics[angle=-90,scale=0.6]{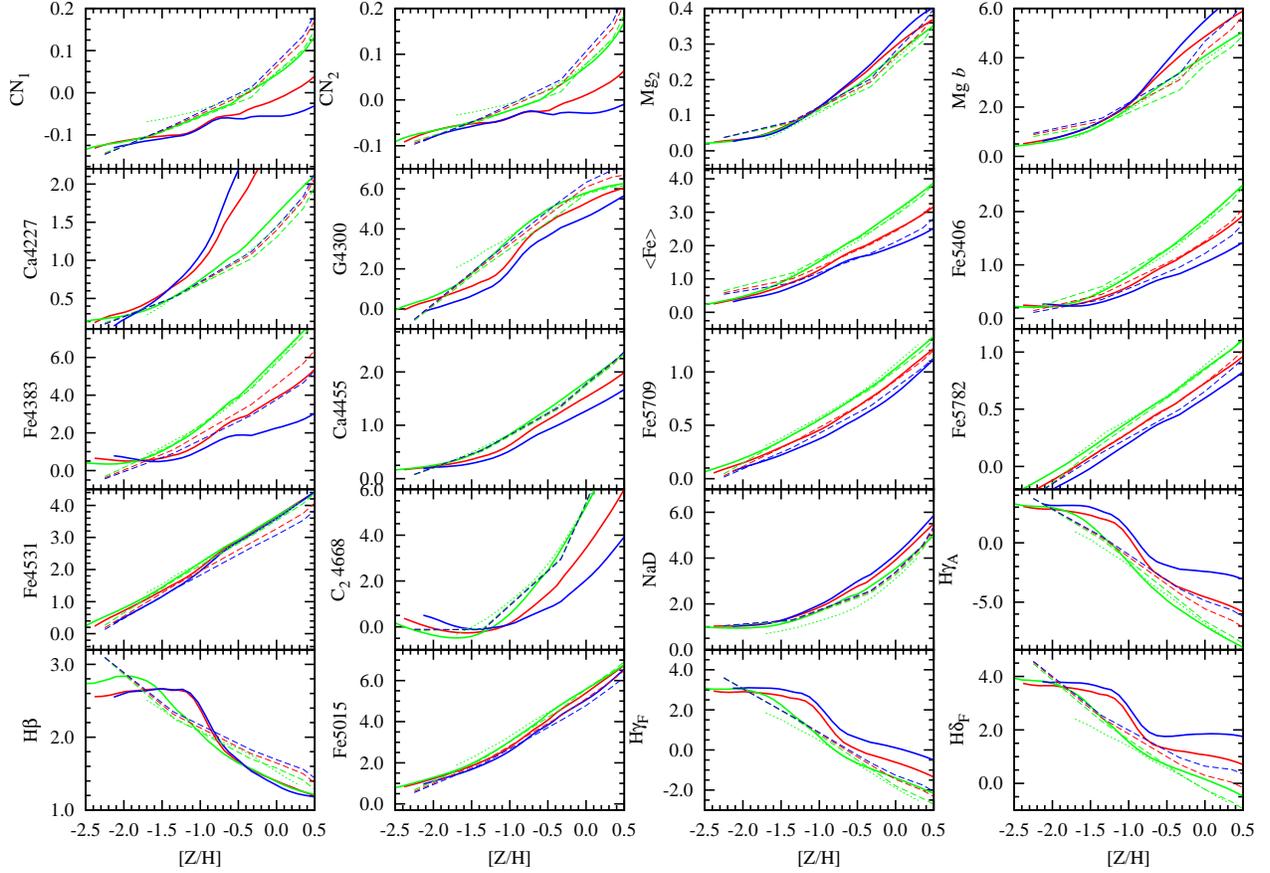}
\caption[]{Comparison of the YEPS models with W94 and TMB03 models. 
All models are the same age of 12~Gyr. 
Solid lines with green, red, and blue colors are predictions of the YEPS model for [$\alpha$/Fe] = 0.0, 0.3, and 0.6, respectively. 
Green, red, and blue dashed lines represent TMB03 models for [$\alpha$/Fe] = 0.0, 0.3, and 0.5. 
Dotted lines in green indicate W94 models with scaled solar abundance.}
\label{1.7}
\end{figure*}

In this Section, we further demonstrate the characteristics of the YEPS model 
by comparing it with other EPS models (the W94 and TMB03 models). 
In order to compare the three different models simultaneously, we use the same metallicity scale $[{\rm Z/H}]$, which is defined as $[{\rm Z/H}] \equiv \log ({\rm Z}/{\rm Z_\odot}) - \log ({\rm H}/{\rm H_\odot})$, where ${\rm Z_\odot}$ and ${\rm H_\odot}$ are respectively the total metallicity and Hydrogen content of Sun. 
Figure~\ref{1.7} displays YEPS (solid lines), TMB03 (dashed lines), and W94 (dotted lines) models.
Although the three models employ heterogeneous sets of stellar evolutionary tracks and EPS modeling schemes, 
they agree well with one another, except for indices CN$_{1}$, CN$_{2}$, G4300, H$\beta$, H$\gamma$, and H$\delta$. 
As discussed in previous Sections, the YEPS model shows different features in these indices 
because the indices are particularly sensitive to blue HB stars. 

Comparison of the YEPS and TMB03 models for the $\alpha$-element-enhanced cases shows 
that the indices insensitive to blue HBs are generally in good agreement with each other. 
In particular, Fe4531, Fe5015, $\left<\rm Fe\right>$, Fe5406, Fe5709, and Fe5782 agree well. 
However, the two models do not match for CN$_1$, CN$_2$, Ca4227, G4300, Fe4383, Ca4455, and C$_2$4668\footnote{Comparison of the YEPS model with the most recent model of \citet{2009ApJ...694..902L} shows good matches for these indices. When the age, [$\alpha$/Fe], and Z of model are assumed to be 12 Gyr, 0.3, and 0.018, respectively, compared to the scaled solar model, the index changes ($\Delta I$) of CN$_1$, CN$_2$, Ca4227, G4300, Fe4383, Ca4455, and C$_2$ 4668 are, respectively,are -0.059, -0.061, 1.01, -0.52, -2.07, -0.28 and -1.84 for the YEPS model, and -0.054, -0.059, 0.43, -0.78, -2.37, -0.35, and -2.57 for the model of \citet{2009ApJ...694..902L}.}
The main reasons for the difference are as follows: 
First, the two models use the different prescription of $\alpha$-elements. 
The TMB03 model uses fixed C, N, and Ca with enhanced $\alpha$-elements (O, Mg, Na, Ne, S, Si, and Ti) 
and depressed iron-peak elements (Cr, Mn, Fe, Co, Ni, Cu, and Zn). 
The YEPS model, on the other hand, adopts the enhanced ratios of $\alpha$-elements from \citet{2000ApJ...532..430V}, 
i.e., fixed C and N, enhanced O, Ne, Na, Mg, Si, P, S, Cl, Ar, Ca, and Ti, 
and depressed Ar and Mn . 
The strong Ca4227 of the YEPS, for instance, can be explained 
by the way we treat Ca as a member of the enhanced $\alpha$-elements group.
Second, TMB03 is based on the scaled-solar abundance isochrones \citep{1997A&A...317..108C,1997ApJ...489..822B,2000A&A...361.1023S}. The model incorporates the effect of $\alpha$-elements enhancement by depressing iron-peak elements (e.g., Fe and Cr) in a way that satisfies the scaling relation ${\rm [Z/H]=[Fe/H]+0.94[\alpha/Fe]}$. 
The effect of the depressed iron-peak elements is not enough to mimic the effect of stellar evolution with a depressed iron-peak elements.
The YEPS model, on the other hand, uses stellar libraries with enhanced $\alpha$-elements, naturally reproducing the effect of depressed iron-peak elements on iron absorption indices without adjusting the iron-peak elements in the stellar atmosphere.

\subsection{Comparison with Observations: Globular Clusters in the Milky Way and M31}

\begin{figure*}

\includegraphics[angle=-90,scale=0.85]{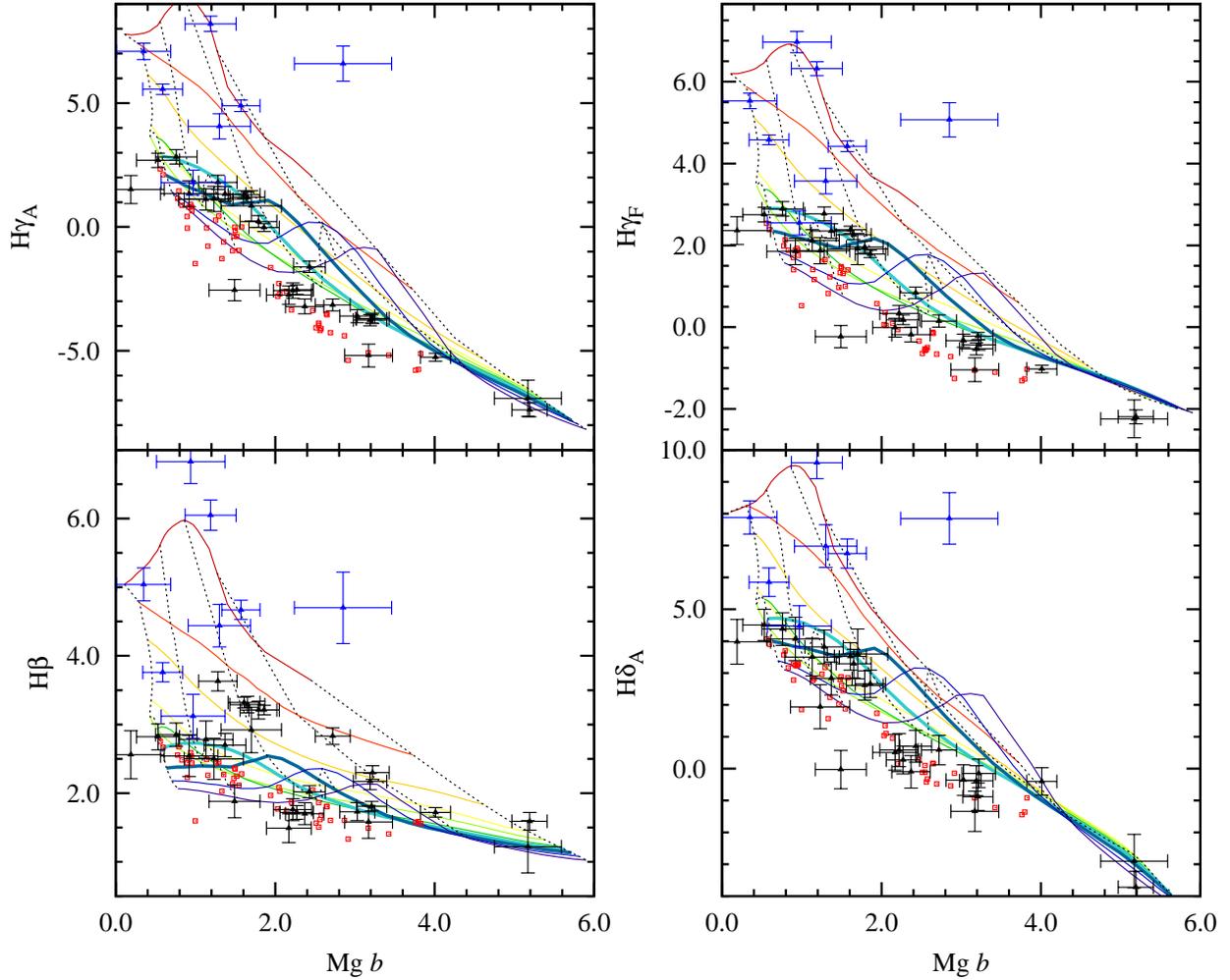}
\caption[]
{Comparisons of the YEPS model (${\rm [\alpha/Fe]}=0.15$) with GCs in the MW and M31. 
Absorption indices of H$\beta$, H$\gamma_A$, H$\gamma_F$, and H$\delta_A$ are presented in this figure.
Solid lines are the same age SSP model and dotted black lines indicate iso-metallicities from [Fe/H] = $-2.5$ to 0.5.
Each solid line corresponds to the age between 1 to 15~Gyr in steps of 1~Gyr from red to blue colors.
Thick lines with turquoise and dark blue colors are the YEPS model for the age of 12 and 13~Gyr, respectively. 
Red squares are GCs in the MW observed by \citet{2005ApJS..160..163S, 2012AJ....143...14S}. 
Black and blue triangles are GCs in M31 from \citet{Beas04}.
Blue triangles are young cluster candidates of M31 based on \citet{Beas04}.
A little offset between metal rich GCs and the model predictions for high order Balmer indices indicates the [$\alpha$/Fe] bias between metal-poor and metal-rich GCs in the MW and M31. 
}
\label{1.8}
\end{figure*}

\begin{figure*}
\includegraphics[angle=-90,scale=0.85]{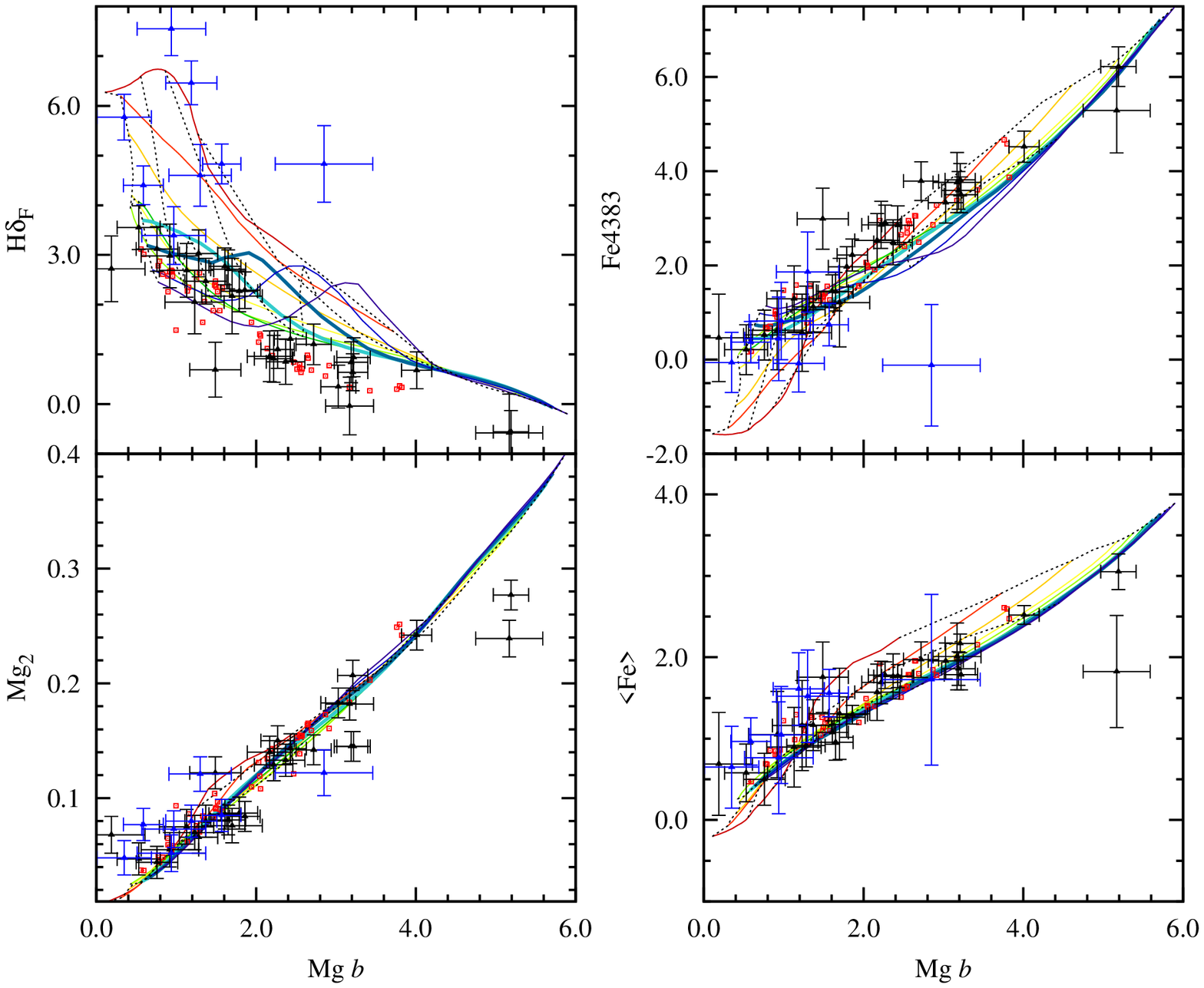}
\caption[]{Same as Figure~\ref{1.8} but for absorption indices of H$\delta_F$, Mg$_2$, $\langle$Fe$\rangle$, and Fe4383.}
\label{1.9}
\end{figure*}

\begin{figure*}
\includegraphics[angle=-90,scale=0.85]{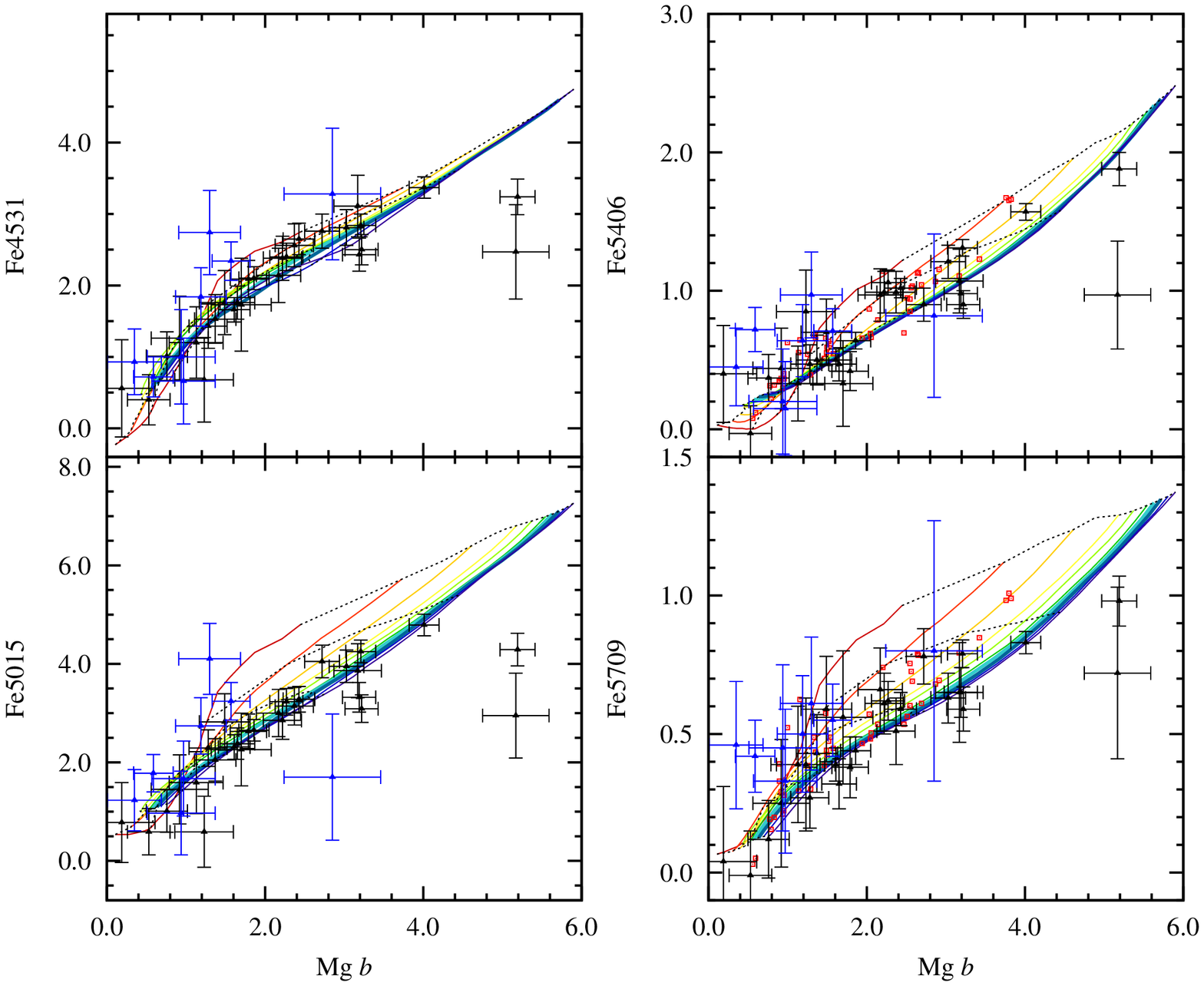}
\caption[]{Same as Figure~\ref{1.8} but for absorption indices of Fe4531, Fe5015, Fe5406, and Fe5709.}
\label{1.10}
\end{figure*}

\begin{figure*}
\includegraphics[angle=-90,scale=0.85]{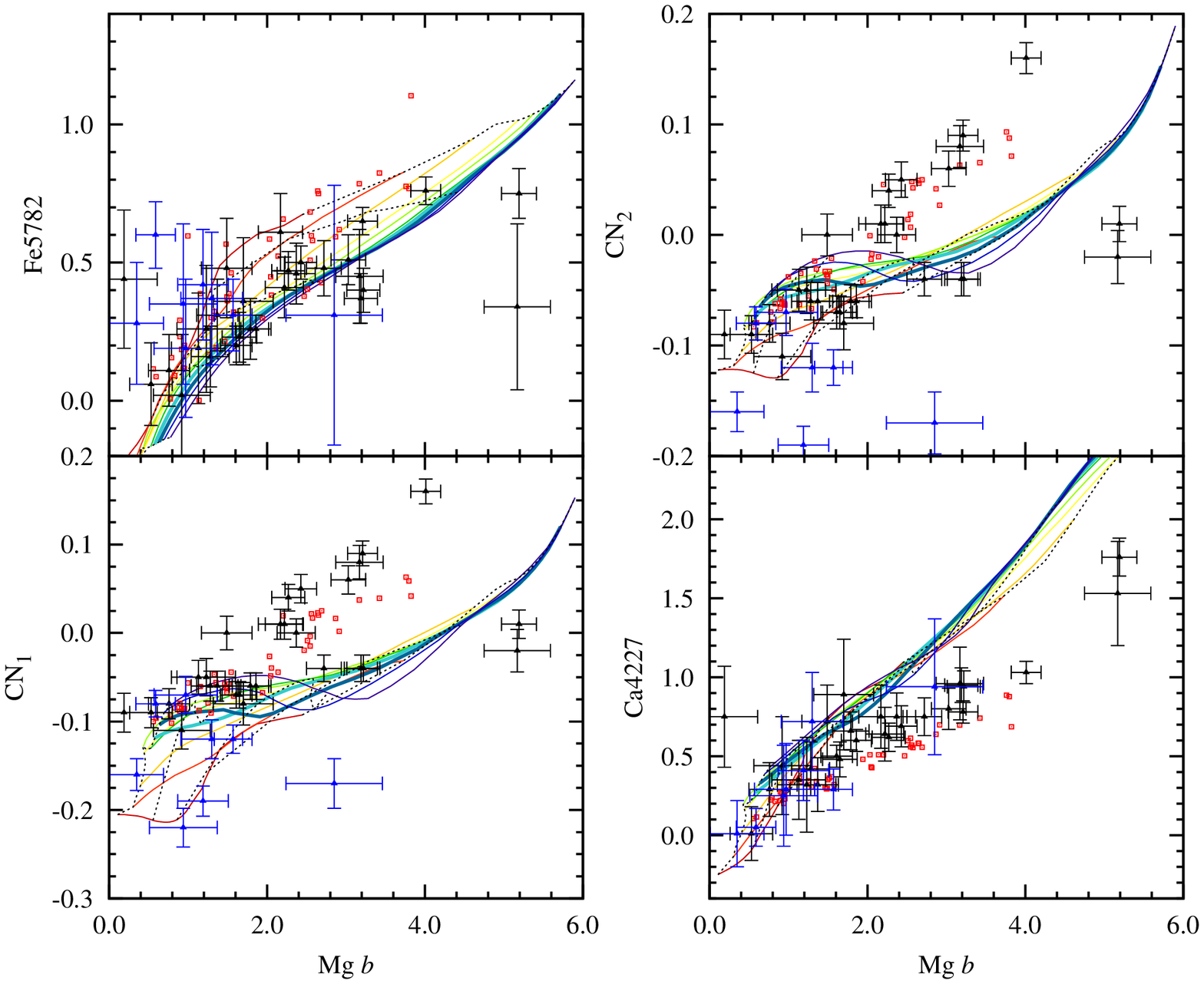}
\caption[]{Same as Figure~\ref{1.8} but for absorption indices of Fe5782, CN$_1$, CN$_2$, and Ca4227.}
\label{1.11}
\end{figure*}

\begin{figure*}
\includegraphics[angle=-90,scale=0.85]{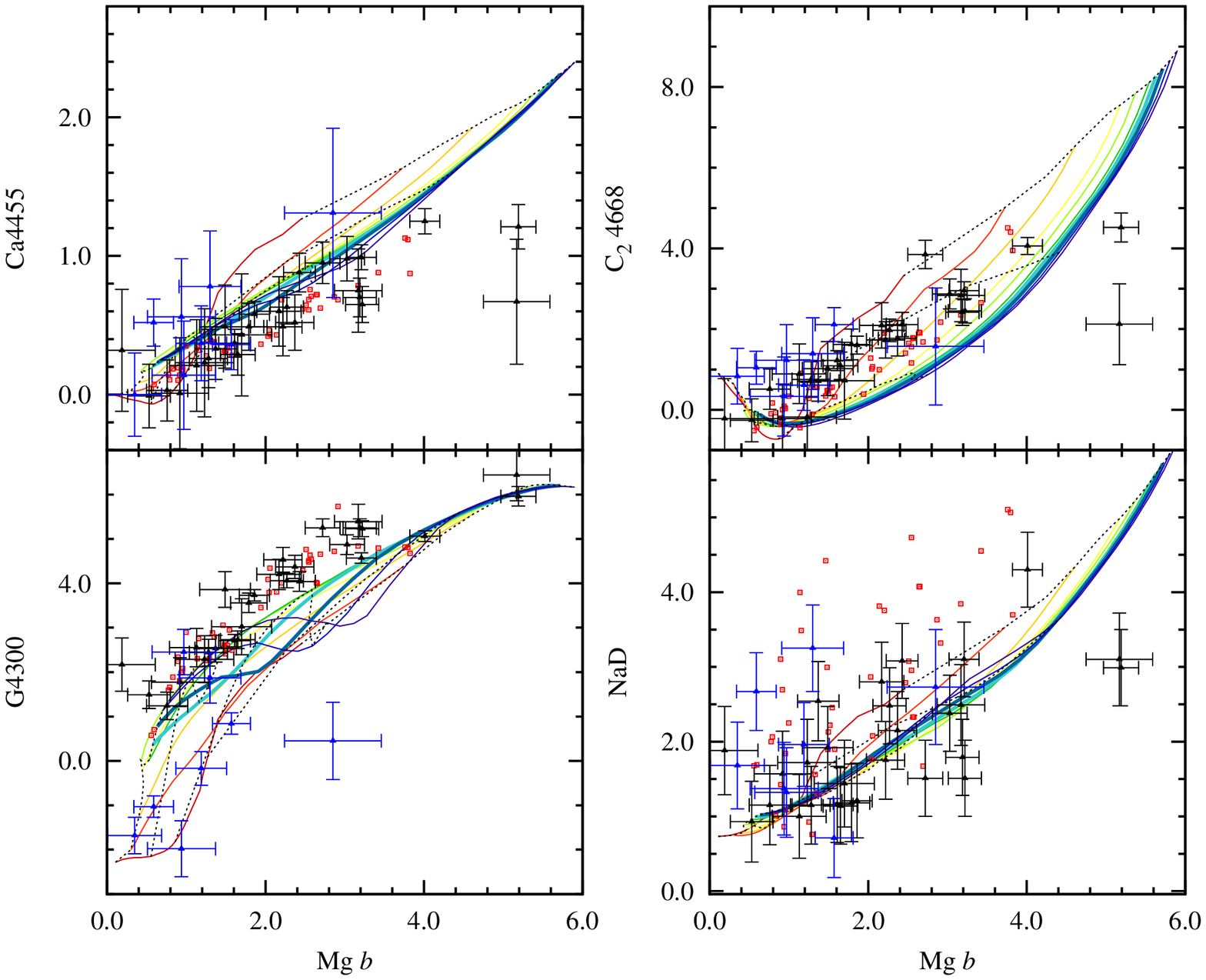}
\caption[]{Same as Figure~\ref{1.8} but for absorption indices of Ca4455, G4300, C$_2$4668, and NaD.}
\label{1.12}
\end{figure*}

\begin{figure*}
\includegraphics[angle=-90,scale=0.85]{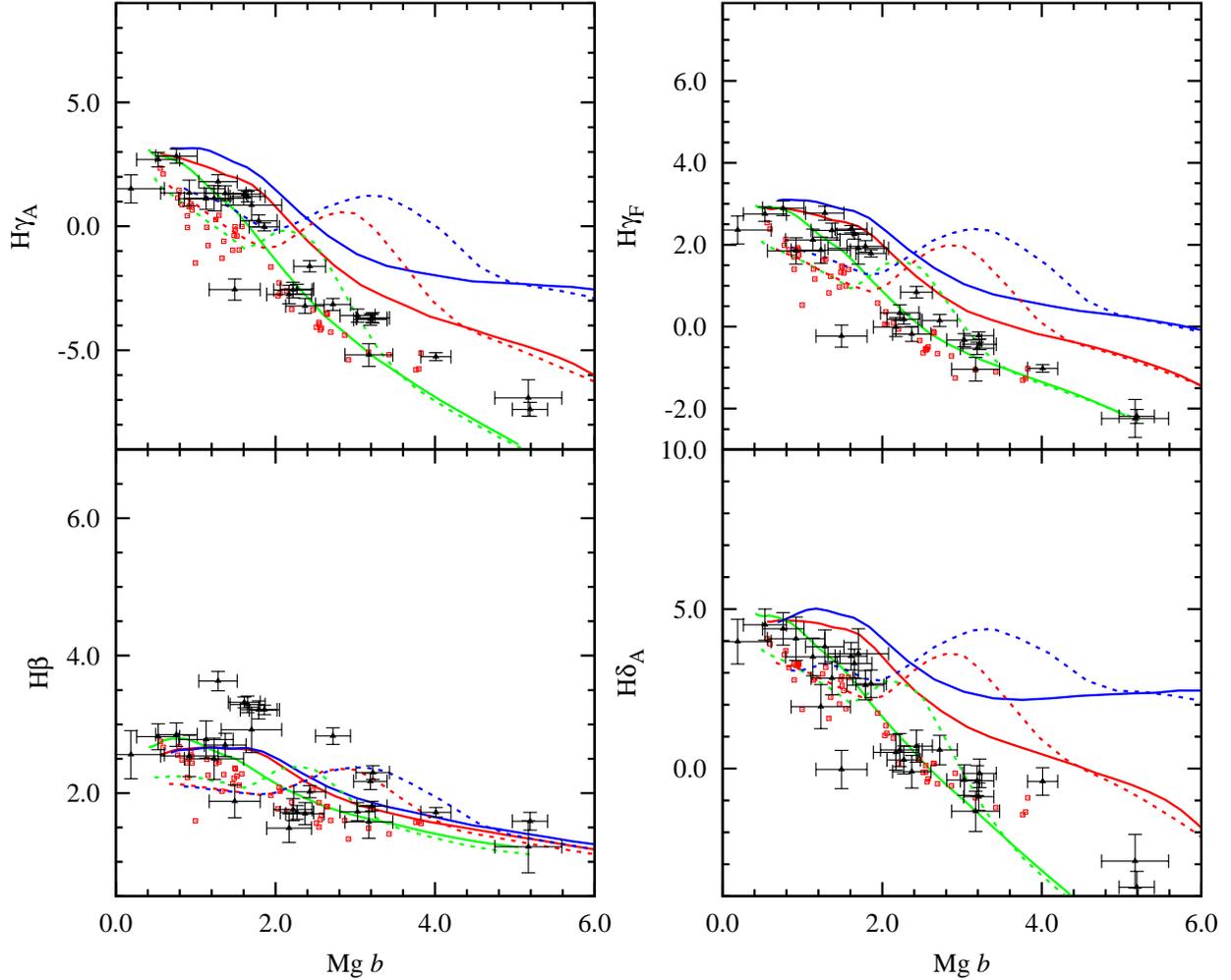}
\caption[]{Comparisons between the YEPS model with enhanced $\alpha$-elements and GCs in the MW and M31.
Absorption indices of H$\beta$, H$\gamma_A$, H$\gamma_F$, and H$\delta_A$ are displayed in this figure.
Green, red, and blue lines represent the YEPS model for [$\alpha$/Fe] = 0.0, 0.3 and 0.6, respectively. 
Solid and dashed lines are the YEPS models for 12 and 14 Gyr, respectively. 
Observed GCs are from the same data used in Figures \ref{1.8} to \ref{1.12} but we have excluded young M31 GC candidates (blue triangles in Figures from \ref{1.8} to \ref{1.12}) for the fair comparison with old age ($>$~12~Gyr) models.
}
\label{1.13}
\end{figure*}

\begin{figure*}
\includegraphics[angle=-90,scale=0.85]{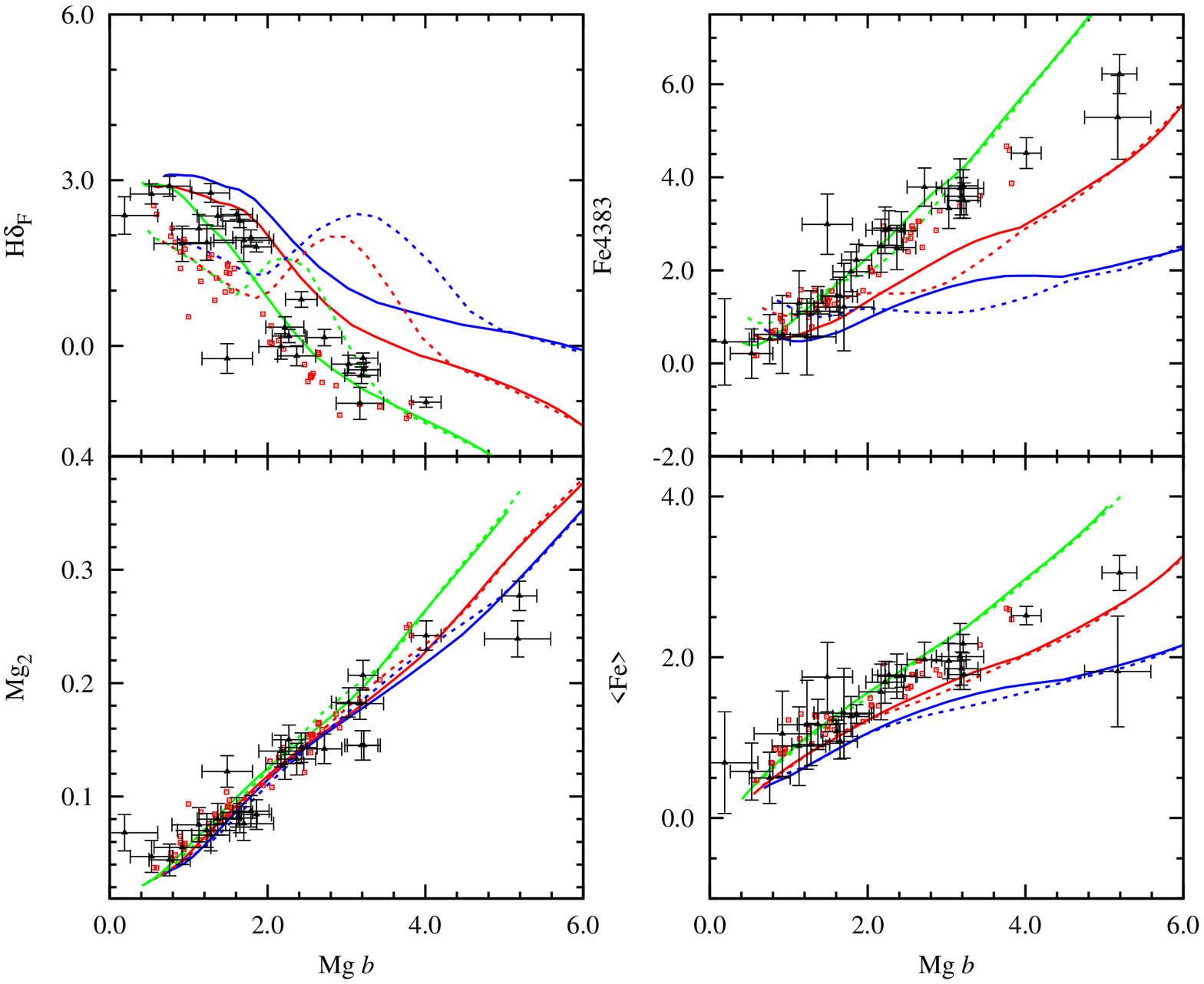}
\caption[]{Same as Figure~\ref{1.13} but for absorption indices of H$\delta_F$, Mg$_2$, $\langle$Fe$\rangle$, and Fe4383.}
\label{1.14}
\end{figure*}

\begin{figure*}
\includegraphics[angle=-90,scale=0.85]{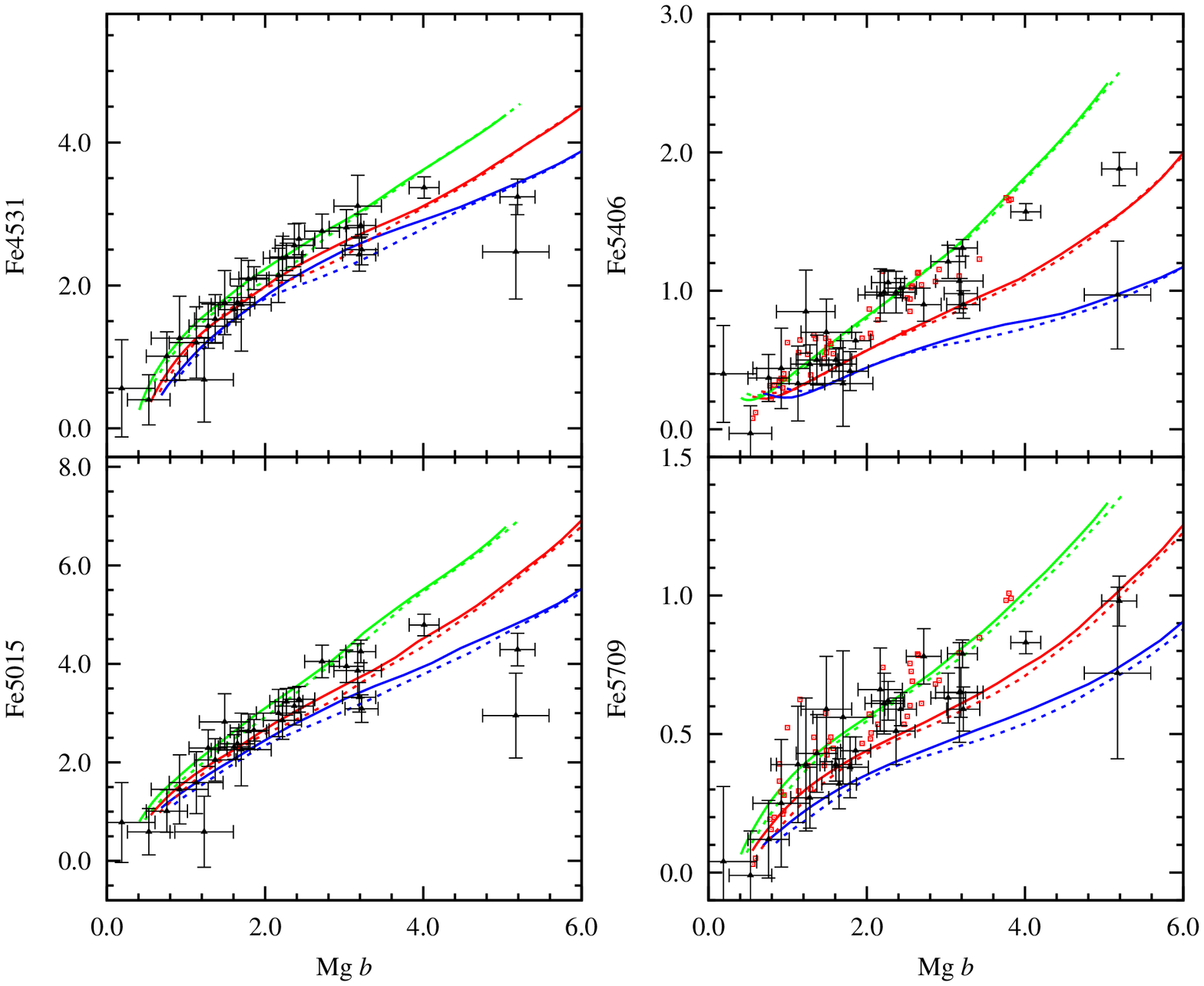}
\caption[]{Same as Figure~\ref{1.13} but for absorption indices of Fe4531, Fe5015, Fe5406, and Fe5709.}
\label{1.15}
\end{figure*}

\begin{figure*}
\includegraphics[angle=-90,scale=0.85]{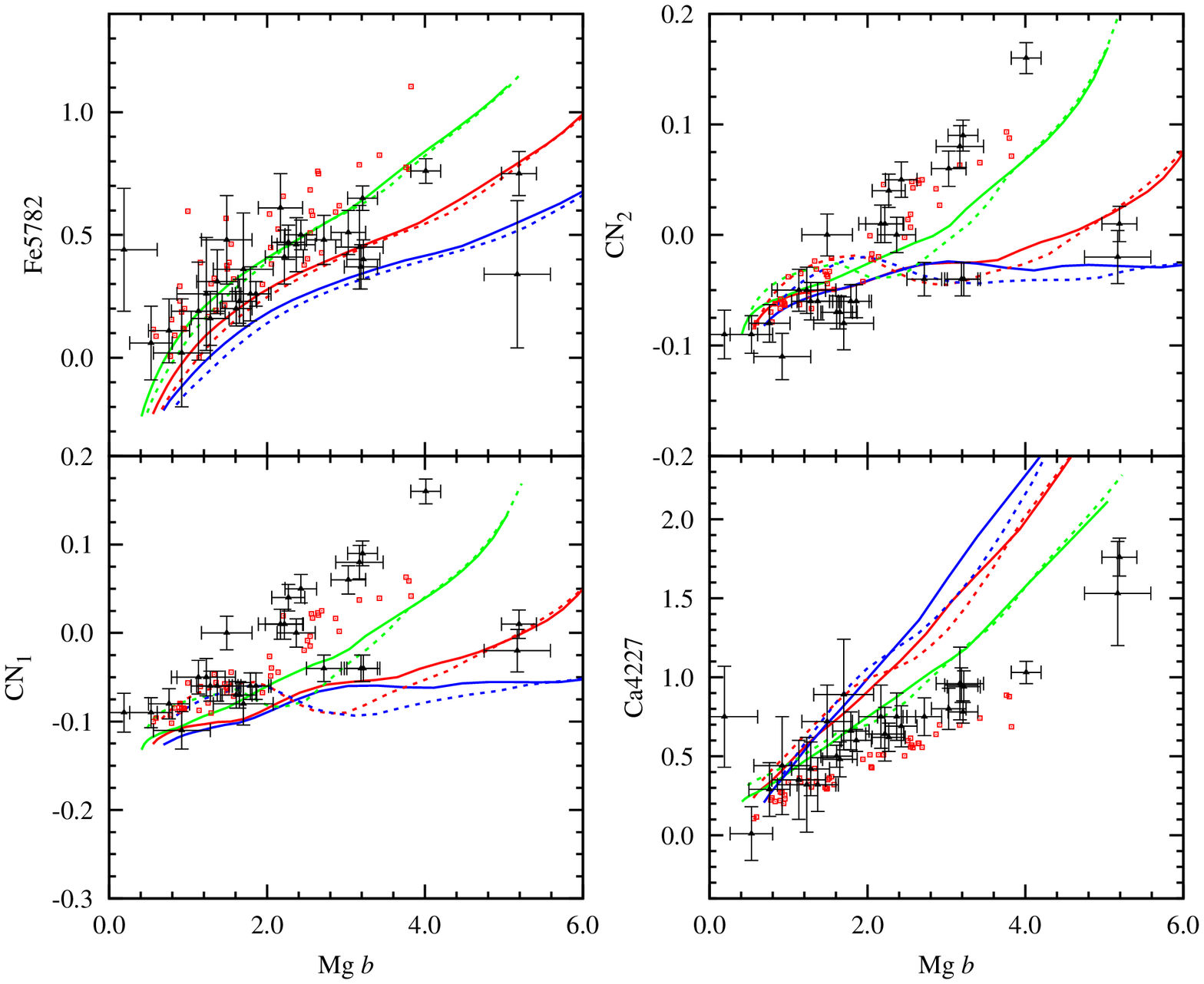}
\caption[]{Same as Figure~\ref{1.13} but for absorption indices of Fe5782, CN$_1$, CN$_2$, and Ca4227.}
\label{1.16}
\end{figure*}

\begin{figure*}
\includegraphics[angle=-90,scale=0.85]{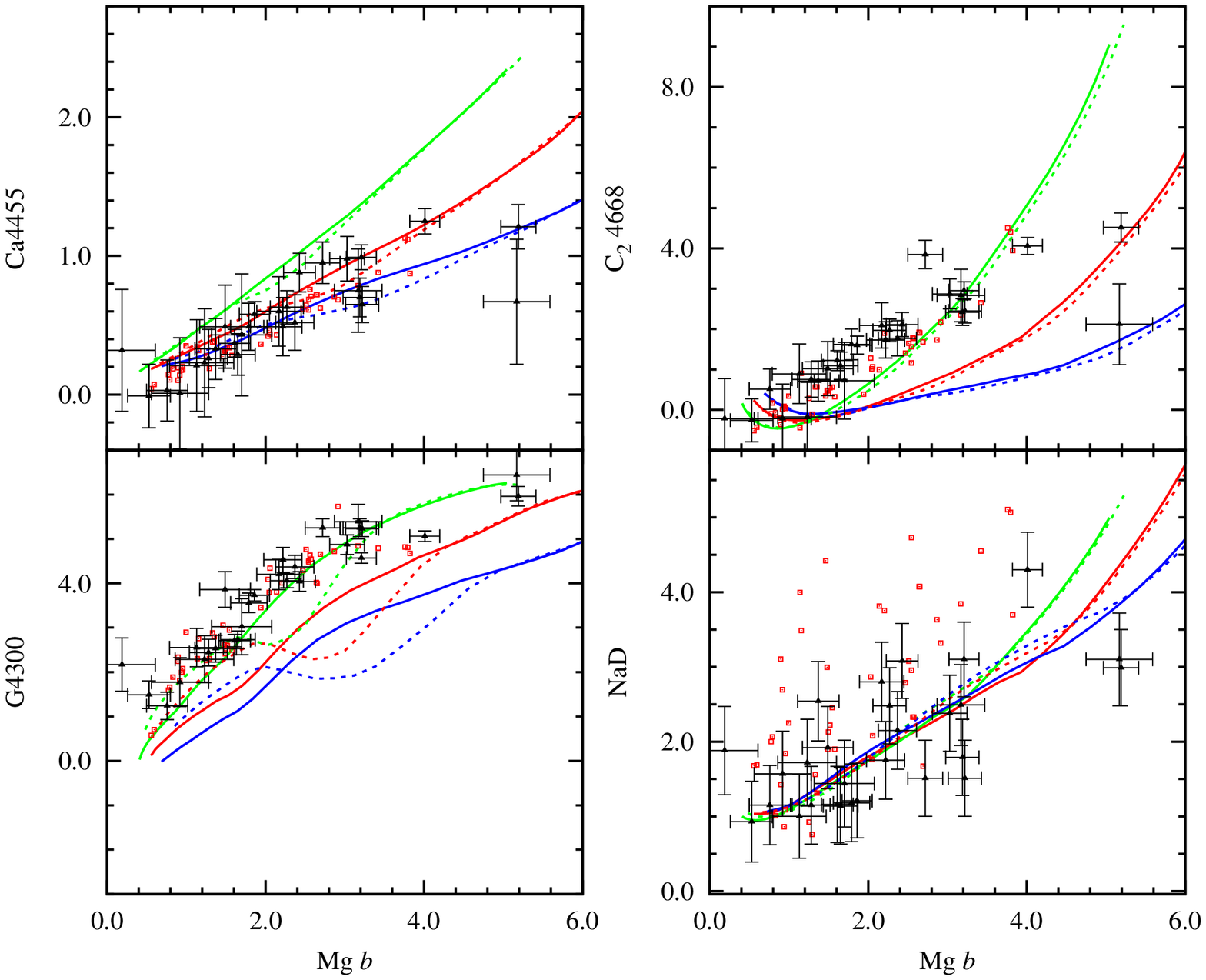}
\caption[]{Same as Figure~\ref{1.13} but for absorption indices of Ca4455, G4300, C$_2$4668, and NaD.}
\label{1.17}
\end{figure*}

Figures \ref{1.8} to \ref{1.17} present the comparison of the YEPS model with the observed data on the GCs in the MW and M31.
The model grids in Figures \ref{1.8} to \ref{1.12} are for various ages (solid lines) and metallicities (dotted lines).
We choose [$\alpha$/Fe] = 0.15\footnote{
Recent spectroscopy of individual stars in the MWGCs confirms that the second generation population in most clusters are depleted in oxygen \citep{2004ARA&A..42..385G, 2005A&A...433..597C, 2009A&A...505..117C}. We can expect a similar chemical inhomogeneity in M31 GCs if the MWGCs are the local counterparts of extragalactic GCs. However, in our models for enhanced $\alpha$-elements, the abundant element oxygen is included in the enhanced $\alpha$-element group. This would explain why GCs in the MW show better fit with models having apparently lower [$\alpha$/Fe] values compared to the [$\alpha$/Fe] measurement based on the [Ca/Fe], [Si/Fe], and [Ti/Fe] of red giant stars in the cluster \citep{2005AJ....130.2140P}.} to consider the [$\alpha$/Fe] distribution for the MWGCs \citep{2003A&A...400..823M, 2007MNRAS.379.1618M, 2010ApJ...708.1335W} and the observed mean [$\alpha$/Fe] of M31 GCs ([$\alpha$/Fe] = 0.14 $\pm$ 0.04; \citealt{2005A&A...434..909P}).
Blue triangles represent young ($\lesssim$ 1 Gyr) GC candidates in M31 \citep{Beas04}.
Figures \ref{1.13} to \ref{1.17} exhibit the line strengths as a function of [Fe/H] for given ages but for different $\alpha$-elements abundance ([$\alpha$/Fe] = 0.0, 0.3 and 0.6).
Data for the MW and M31 GCs are obtained from \citet{2005ApJS..160..163S, 2012AJ....143...14S} and \citet{Beas04}, respectively. 
For the fair comparison with old age ($>$ 12 Gyr) models, we have taken out from young M31 GC candidates \citep{Beas04} in these Figures.
With a few exceptions, the old GCs in MW and M31 populate well along the 12-Gyr model for the most indices.
The theoretical predictions of the YEPS model for SSPs at 12~Gyr are given in Tables~\ref{tab.1.5}--\ref{tab.1.7}.

Notes on individual indices shown in Figures \ref{1.8} to \ref{1.17} are as follows.

\textbullet\ H$\beta$, H$\gamma_{A}$, H$\gamma_{F}$, H$\delta_{A}$, and H$\delta_{F}$ ---
Overall shapes of model Balmer lines are in good agreement with the observation in Figures~\ref{1.8}, \ref{1.9}, \ref{1.13}, and \ref{1.14}.
A little offset between metal-rich GCs and the model predictions for high order Balmer indices (Figure~\ref{1.8} and \ref{1.9}) indicates the [$\alpha$/Fe] bias between metal-poor and metal-rich GCs in the MW and M31 (see Figure~\ref{1.13} and \ref{1.14}).
In the metal-poor regime, the 12-Gyr model GCs contain hot ($>$\,8,000\,K) HB stars that lead to stronger Balmer lines than the model without HB stars. 
The extensive study of GCs in M31 by \citet{2011AJ....141...61C} confirms that Balmer indices of GCs with blue HBs (see their Figure~10) are on average stronger by $\Delta$H$\beta \simeq 0.6$\,{\small \AA} and $\Delta$H$\delta_F \simeq 2.0$\,{\small \AA}.
As a consequence, the model line is highly inflected, reproducing the observed behavior of the index-index relations. 

The effect of $\alpha$-elements on Balmer indices differs between H$\beta$ and higher-order Balmer indices. 
Figure~\ref{1.13} shows that H$\beta$ is hardly affected by any $\alpha$-elements.
In contrast, H$\delta$ and H$\gamma$ are sensitive to C, Mg, Fe, and total metallicity (Figures~\ref{1.13} and \ref{1.14}). 
The H$\delta_{A}$ index is the most sensitive to $\alpha$-elements enhancement in the metal-rich regime because of its sensitivity to Fe abundance of giant stars (K05). 
Note that the effect of HB stars on the Balmer indices is universal regardless of $\alpha$-elements enhancement.

\textbullet\ Mg$_{1}$, Mg$_{2}$, and Mg\,$b$ ---
The YEPS model well reproduces the observations of Mg$_{1}$, Mg$_{2}$, and Mg\,$b$. 
The lines are simultaneous affected by $\alpha$-elements, 
[$\alpha$/Fe] does not change the overall shape of, for instance, the Mg\,$b$--Mg$_{2}$ relation (Figure~\ref{1.14}).

\textbullet\ Fe5270, Fe5335, and $\langle$Fe$\rangle$ ---
The models and observations show a remarkable agreement in Figure~\ref{1.9}. 
The lines are widely used as indicators of iron abundance, 
yet are fairly sensitivity to [$\alpha$/Fe] (Figure~\ref{1.14}).

\textbullet\ Fe4383 ---
The YEPS model well reproduces Fe4383 for GCs both in the MW and M31 (Figure \ref{1.9}). 
Since Fe4383 traces Fe and the total metallicity simultaneously, 
the Fe4383 model is very sensitive to variations of [$\alpha$/Fe] (Figure \ref{1.14}).

\textbullet\ Fe4531 ---
Our model of Fe4531 agrees well with M31 GCs (Figure~\ref{1.10}). 
This index is a good iron abundance indicator that is relatively less sensitive to [$\alpha$/Fe] (Figure~\ref{1.15}).

\textbullet\ Fe5015 ---
The YEPS model shows a good agreement with the M31 GCs (Figure \ref{1.10}). 
The Fe5015 strength is generally involved with the Ti and Mg abundance 
in low MS stars and giant stars, as well as total metallicity near TOs. 
Yet, the effect of [$\alpha$/Fe] on Fe5015 is only modest (Figure \ref{1.15}).

\textbullet\ Fe5406, Fe5709, and Fe5782 --- GCs in the MW and M31 show good matches with the models. 
The slight offset of MWGCs in Fe5782 (Figure \ref{1.11}) 
is likely due to the well-known problem of miscalibration of fitting functions (TMB03). 
The $\alpha$-element sensitivity of these indices are 
similar to those of Fe5270 and Fe5335 (Figures~\ref{1.15} and \ref{1.16}).

\textbullet\ CN$_{1}$ and CN$_{2}$ ---
The models for CN$_{1}$ and CN$_{2}$ do not show good fits to the observational data in Figure \ref{1.11}. 
Compared to observational data, our models predict on average 0.1 mag lower values in the metal-rich regime. 
The inferred [$\alpha$/Fe] seems unreasonably high 
and likely due to the poor calibration of fitting functions (TMB03; \citealt{Lee05}).
{The poor calibration of fitting functions can be explained by the effect of CNONa anticorrelation found in the MWGCs with the second generation stellar population \citep{2011ApJ...734...72C}.}
CN$_{1}$ and CN$_{2}$ indices, even for the $\alpha$-element enhanced models, 
are affected fairly by blue HB stars (see solid and dashed lines in Figure \ref{1.16}).

\textbullet\ Ca4227 ---
Our models for Ca4227 show a small offset with observations.
Observations occupy slightly low EWs compared to the model in Figure \ref{1.11} and Figure \ref{1.16}.
As suggested by TMB03 and \citet{2009AJ....138.1442L}, 
the enhanced C and N model can depress EW of Ca4227 
but we do not consider the case of the enhanced C and N model in this study.
{Note that more realistic model with the effect of CNONa anticorrelation would have decreased strengths of Ca4227 \citep{2011ApJ...734...72C} and this may explain the offset between the YEPS model and observations.}
Since Ca4227 is very sensitive to Ca and C, but is not so to Mg, 
the Ca4227 line deepens with increasing [$\alpha$/Fe] (Figure \ref{1.16}).

\textbullet\ Ca4455 ---
The Ca4455 model for [$\alpha$/Fe] = 0.15 predicts on average 0.3\,{\small \AA} higher EWs 
compared to the observations (Figure \ref{1.12}),
and the observations are better reproduced by the model with [$\alpha$/Fe] $\simeq$ 0.3 (Figure \ref{1.17}). 
Since Ca4455 is insensitive to the variation of any of elements studied in K05, 
the effect of [$\alpha$/Fe] shown in Figure \ref{1.17} come solely from the variation of Fe 
due to the increased [$\alpha$/Fe] at a given $Z$.
This explains why Ca4455 shows a different response compared to Ca4227 even though we treat Ca as a member of enhanced $\alpha$-elements group (Table~\ref{tab.1.2}).  

\textbullet\ G4300 ---
The YEPS model with enhanced $\alpha$-elements ([$\alpha$/Fe] = 0.15) for the G4300 
shows a little offset from the observations (Figure \ref{1.12}).
Rather, GCs in the MW and M31 follow the model with [$\alpha$/Fe] = 0.0 well (Figure \ref{1.17}).
Given that the other lines agree with the observations, 
our G4300 model seems too sensitive to [$\alpha$/Fe].

\textbullet\ C$_2$4668 ---
The YEPS model for the C$_2$4668 index offsets from the observations (Figure \ref{1.12}). 
The model with enhanced [$\alpha$/Fe] (Figure \ref{1.17}) 
shows great sensitivity to [$\alpha$/Fe].
C$_2$4668 is a measurement of C abundance, and is insensitive to all other $\alpha$-elements.
Because the YEPS model with enhanced [$\alpha$/Fe] used fixed C abundance, the strong sensitivity to the $\alpha$-elements enhancement in Figure \ref{1.17} comes mainly from the [Fe/H] variation due to the $\alpha$-element contents at a given Z.

\textbullet\ NaD ---
The YEPS model for NaD index is weaker than the observations. Part of the reason for this discrepancy is Na absorption by the interstellar medium (TMB03).
{As suggested by \citet{2011ApJ...734...72C}, more realistic model with increased Na abundance caused by the observed CNONa anticorrelation in the second generation population would increase the strength of NaD.}
The Na sensitivity of NaD plays an important role in the $\alpha$-element enhanced model by decreasing the strengths of NaD {at given Mg\,$b$} (Figure \ref{1.17}).

Based on the comparison shown in Figures~\ref{1.8} to \ref{1.17}, we identify several Lick indices that are most appropriate for the estimation of  metallicity, age, and $\alpha$-elements enhancement of stellar populations. 
(1) For determination of iron abundance, Fe4383, Fe4531, Fe5015, Fe5270, and Fe5335 are recommended; (2) For age-dating, Balmer indices such as H$\beta$, H$\gamma_{A}$, H$\gamma_{F}$, H$\delta_{A}$, and H$\delta_{F}$; and (3) For measuring $\alpha$-elements enhancement, Mg$_{1}$, Mg$_{2}$, and Mg\,$b$.

\section{DISCUSSION}

\subsection{Estimation of Age and Metallicity with YEPS}

In this Section, we discuss how the effects of HB stars and $\alpha$-elements enhancement in the model are combined to affect the age and metallicity estimation of SSPs. 
To this aim, we select three typical MWGC , 47 Tucanae (NGC~104), NGC~6284, and M67 (NGC~2682), representing old metal-rich, old metal-poor, and young metal-rich GC populations, respectively. These clusters have well-studied CMDs and show no strong evidence of multiple stellar populations reported by recent studies \citep{Lee99, 1997ApJ...486L.107L, 2009Natur.462..480L, Han09, 2009Natur.462..483F, 2007ApJ...661L..53P, 2009A&A...493..539M}. 

\begin{figure*}
\includegraphics[angle=-90,scale=0.7]{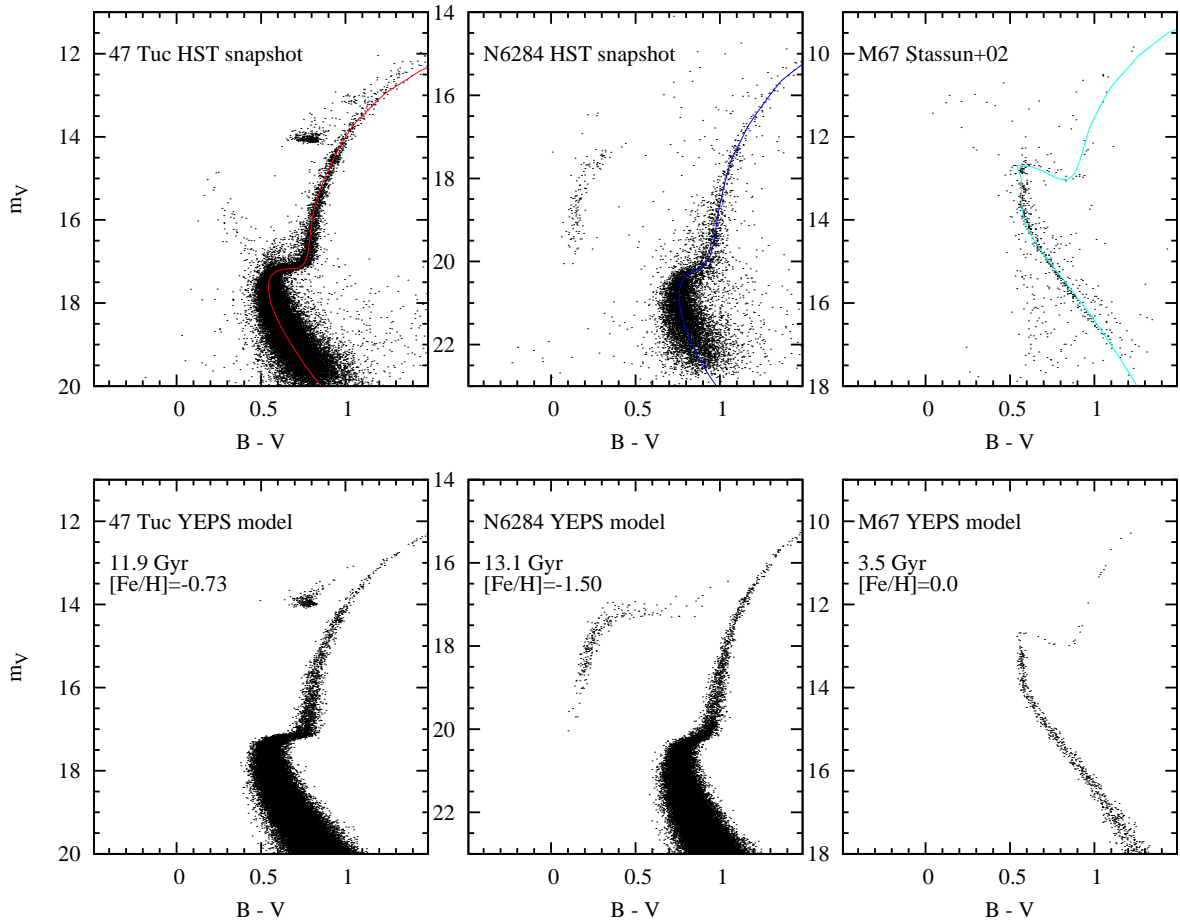}
\caption[]
{
Comparison of the observed 47~Tuc, NGC~6284 \citep{Piot02}, and M67 \citep{Stas02} with the synthetic CMD models (bottom panels). 
Observed CMDs are displayed with matched Y$^2$-isochrones.}
\label{1.18}
\end{figure*}

Figure~\ref{1.18} displays the observed and synthetic CMDs for 47~Tuc, NGC~6284, and M67.
The free parameters used to match the synthetic CMDs with the observed ones are age and metallicity.
To simulate observational errors in our model, we also carried out Monte Carlo simulations based on the actual observational uncertainties.
The best-fit parameters of each model are summarized in Table~\ref{tab.1.4}.

\begin{figure*}
\includegraphics[angle=-90,scale=1.1]{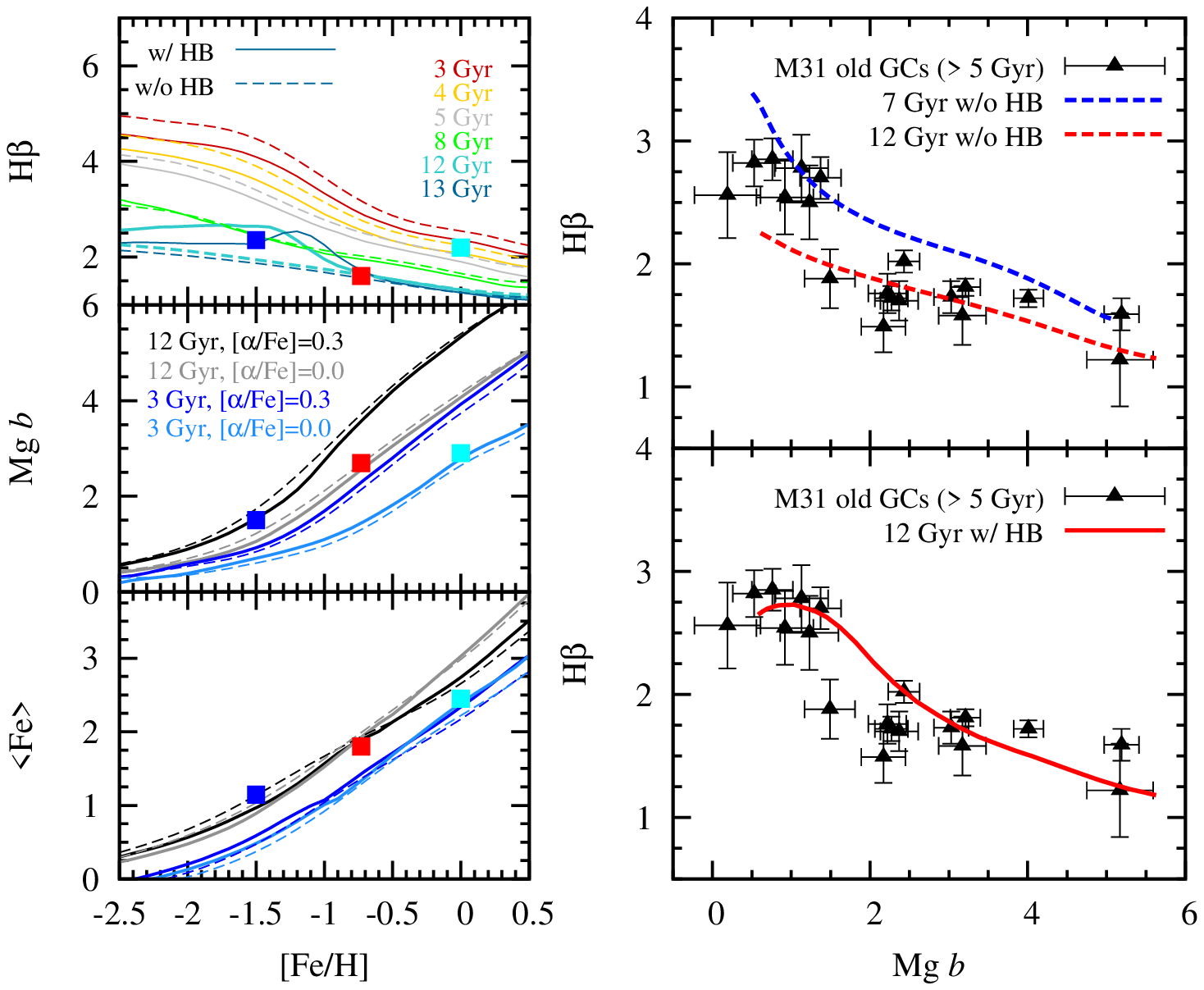}
\caption[]
{Comparison of the YEPS model with the observed absorption indices of NGC~6284, 47~Tuc \citep{2005ApJS..160..163S, 2012AJ....143...14S}, and M67 \citep{Schi07}, as well as GCs in M31 \citep{Beas04}. 
Blue, red, and cyan squares in the left panels are observed as integrated absorption indices of NGC~6284, 47~Tuc, and M67. 
Solid and long dashed lines in all panels are models with and without HB stars.
Red, orange, grey, green, turquoise, and blue lines in the left top panel represent the H$\beta$ models for the ages of 3, 4, 5, 8, 12 and 13 Gyr with [$\alpha$/Fe] = 0.3. 
Grey and black lines in the middle and bottom panels on the left side are models for [$\alpha$/Fe] = 0.0 and 0.3 at the age of 12~Gyr.
Light blue and blue lines in the same panels are models for [$\alpha$/Fe]=0.0 and 0.3 at the age of 3~Gyr.
In the right panels, the effect of HB stars on the age-dating with the H$\beta$ is displayed with M31 GCs \citep{Beas04}.
{Model lines in the right panels are for [$\alpha$/Fe]=0.15.}
} 

\label{1.19}
\end{figure*}

The left column of Figure~\ref{1.19} compares the absorption indices between the observations and models for the typical GCs in the MW.
{Since the YEPS models without HB stars generate very similar results to other models with red clump single-mass HB \citep{Lee00}, we present the YEPS model with and without HB stars in Figure~\ref{1.19} (1) to highlight the effect of HB, and (2) to compare our model with other models with red clump single-mass HBs.}
The top panel shows that the H$\beta$ indices of the GCs are reproduced better by the model with the HB effect and the ages based on the absorption lines are in better agreement with the ages (Table~\ref{tab.1.4}) derived from synthetic CMDs in Figure~\ref{1.18}.
In particular, {the systematic variation of HB morphologies with respect to metallicity and age} is essential to explain the enhanced H$\beta$ strength of NGC~6284---an old, metal-poor system with well-developed blue HB stars.
The age of NGC~6284 would be estimated to be 8--9 Gyr without the HB effect, which is inconsistent with its derived age (13.1 Gyr) from the MSTO and HB morphology in Figure~\ref{1.18}.  
On the other hand, the absorption strengths of 47~Tuc and M67 show reasonable agreement with the YEPS models for 12-Gyr (solid cyan lines) and 3.5-Gyr (between solid red and orange lines) GCs, respectively.
The models for both metal-rich and young clusters possess red HB stars.
For a 3-Gyr model GC, the red HB reduces H$\beta$ by 0.2\,{\small \AA}, 
which corresponds to ${\sim}$\,1 Gyr. 
Therefore, in order to derive accurate ages of stellar populations based on the EPS model prediction of Balmer strengths, one should check first whether the EPS model has well calibrated HBs for both blue-HBs (e.g., NGC 6284) and red-HBs (e.g., 47~Tuc and M67). 

The right column of Figure~\ref{1.19} is similar to the top panel of the left column, 
but shows the age dating of M31 GCs \citep{Beas04}. 
{In these panels, based on the $\alpha$-elements enhancement of GCs in M31 \citep{2005A&A...434..909P}, we have used the model of [$\alpha$/Fe] = 0.15 for the comparison.}
The metal-poor GCs in M31 exhibit stronger H$\beta$ by $\sim$1\,{\small \AA} than the metal-rich counterparts.
The upper panel of the {right} column {shows}, when the effect of HBs is not considered, the models give $\sim$\,5 Gyr younger ages for the metal-poor GCs {compared to} the metal-rich GCs in M31.
With the general trend of age--metallicity relation in mind, it would be difficult to interpret that the mean age of metal-poor GCs is younger than that of metal-rich GCs.
By contrast, the lower panel shows that the systematically enhanced Balmer indices of the metal-poor M31 GCs are well reproduced by our single age, 12-Gyr model GCs with HB stars.
Indeed, \citet{2009A&A...507.1375P} directly detected well developed blue HB stars in the metal-poor GCs in M31 (e.g., B010, B220, B224, and B366) based on the {\it HST}/ACS CMDs.

Back in the left column of Figure~\ref{1.19}, the middle and bottom panels highlight the effects of HBs and $\alpha$-elements enhancement on the metallicity determination of SSPs.
The effects of $\alpha$-elements and HB stars go in the opposite direction on absorption strengths of Mg\,$b$ for given [Fe/H]'s; 
the inclusion of blue HB stars decreases indices, whereas the enhancement of $\alpha$-elements increases indices.
While the HBs exert only a marginal effect on determining [Fe/H], 
the $\alpha$-elements have a marked effect. 
For instance, the effect of enhanced $\alpha$-elements ([$\alpha$/Fe] = 0.3) on Mg\,$b$ is about 10 times greater than that of HBs when the age and [Fe/H] are assumed to be 12 Gyr and 0.0 dex, respectively.
Note that the model line of Mg\,$b$ for 12~Gyr GCs with [$\alpha$/Fe] = 0.3 is almost identical to that for 4~Gyr GCs with [$\alpha$/Fe] = 0.0 due to strong sensitivity of the Mg\,$b$ to the $\alpha$-elements.
In order to determine accurate [$\alpha$/Fe], therefore, the age dating and the metallicity determination of SSPs should be carried out at the same time.
The $\langle$Fe$\rangle$ index, on the other hand, is less sensitive to [$\alpha$/Fe] than Mg\,$b$, 
which makes $\langle$Fe$\rangle$ a more accurate [Fe/H] indicator for the SSPs 
than other indices sensitive to [$\alpha$/Fe] (e.g., Mg\,$b$ and Mg$_{2}$).

Since the age dating with Balmer indices is usually applied to the samples of elliptical galaxies 
\citep[e.g.,][]{Trag00, 2003AJ....125.2891C, 2005ApJ...621..673T, 2005ApJ...632..137N, 2008MNRAS.386..715T, 2009ApJ...693..486G}, 
the effect of HB stars is also important for composite stellar populations.
The effect of HB stars on the integrated observables of giant elliptical galaxies 
should be limited because their mean metallicities are as high as [Z/H] $>$ 0.0, 
for which hot HB stars are rare.
But the metallicity spread of ellipticals, nevertheless, allows them 
to have a certain fraction of low-metallicity stars and thus hot HB stars accordingly \citep{Park97, 2011ApJ...740L..45C}.
Moreover, the effect of hot HB stars in dwarf elliptical galaxies is more important 
because the metallicity of dwarf ellipticals are generally lower than that of giant ellipticals.
In the context of galaxy downsizing \citep[e.g.,][]{1996AJ....112..839C}, 
the age dating of giant and dwarf elliptical galaxies with well calibrated HB models 
is crucial issue for determining the formation history of elliptical galaxies.
We will fully discuss this issue in our forthcoming paper (C. Chung et al. 2012, in preparation).

\subsection{Distributions of Absorption Indices in Extragalactic GCs}

Ever since the recognition of bimodal broadband color distributions of GCs in massive early-type galaxies 
\citep{Zepf93, 1996AJ....111.1529G, 1997AJ....113..887F, Gebh99, Kund01, Lars01, Jord02, 2004A&A...415..123P, West04, 2006AJ....132.2333S, 2006ApJ...639...95P, 2006ApJ...636...90H, 2009ApJS..180...54J, 2011ApJ...728..116L}, 
the phenomenon has been interpreted as the presence of two GC subsystems within individual galaxies. 
Three major ideas have been put forward to explain it, 
including the merger model (e.g., \citealt{Ashm92}), the in-situ (e.g., \citealt{Forb98}) and accretion (e.g., \citealt{Cote98}) scenarios.

More recently, however, \citet{Yoon06} and \citet{2011ApJ...743..149Y, 2011ApJ...743..150Y} 
suggest an alternative explanation for the GC color bimodality that does not necessarily invoke two GC subpopulations.
Yoon et al. show that the theoretical metallicity--color relations are inflected 
and that such relations can generate bimodal color distributions from broad metallicity spreads, even if they are unimodal.
The \citet{Yoon06} model of GC colors indicates that the HB effect is the most important for the inflection on the metallicity--color relations. HBs also have strong effect on the absorption index versus metallicity relations (IMRs), 
and thus one can put the Yoon et al. explanation to the test by examining the metallicity--index nonlinearity and the resulting index distributions.

\subsubsection{Conversions from Metallicity Spreads to Index Distributions}

\begin{figure*}
\includegraphics[angle=0,scale=0.72]{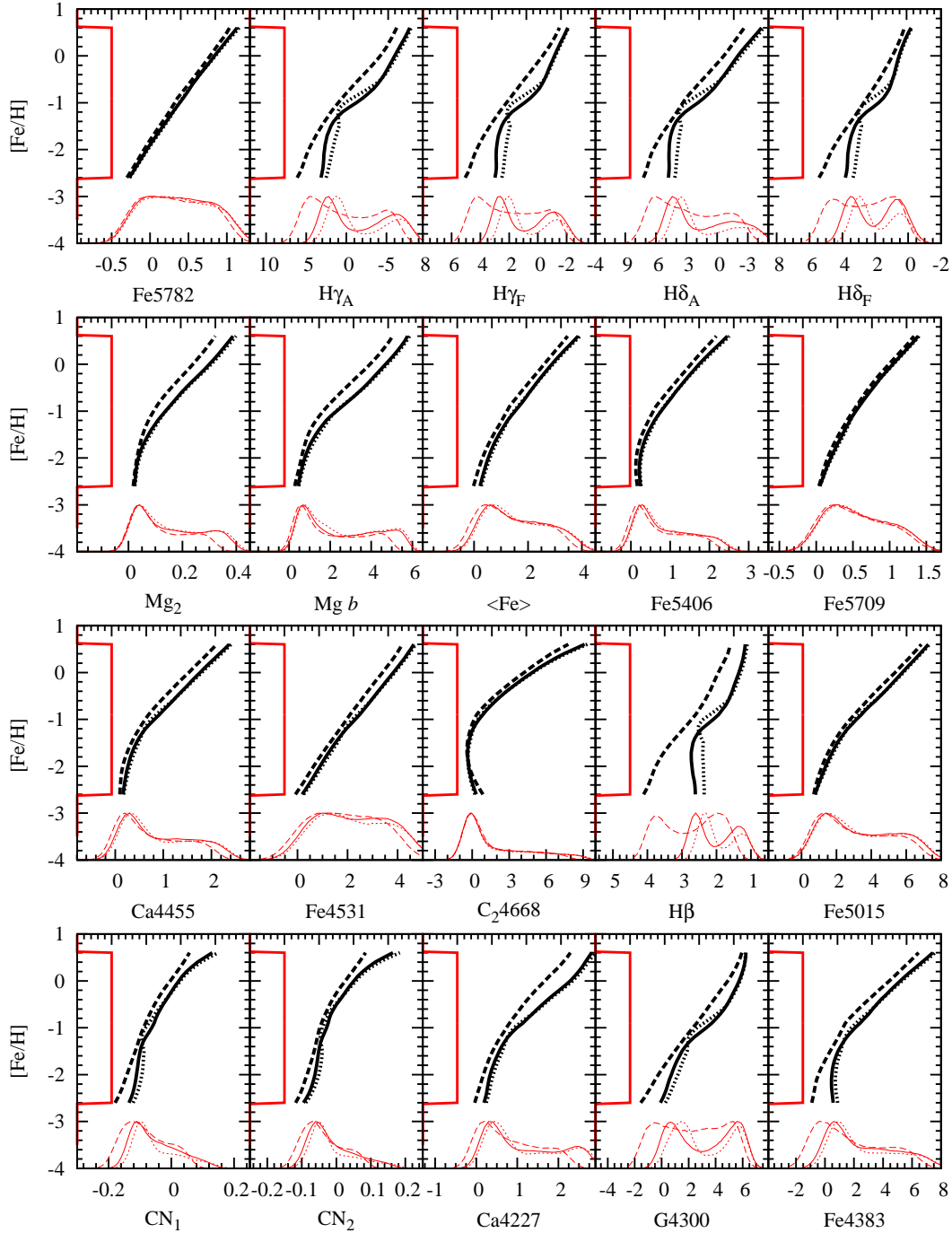}
\caption[]
{\small The test of the projection effect with the box-shaped metallicity distribution function (MDF) on various Lick absorption indices. 
{The box-shaped MDFs of 10$^6$ model GCs are shown along with y-axis.}
Black dashed, solid and dotted lines in each plot indicate 5, 12, and 13~Gyr models with ${\rm [\alpha/Fe] = 0.15}$ for given Lick indices. 
Red dashed, solid, and dotted lines in the bottom of each plot show the result of box-MDF projection to various Lick absorption indices. 
Absorption index distributions of each line type are corresponding to the projection through the models with the same line types. 
}
\label{1.20}
\end{figure*}

Figure~\ref{1.20} shows the conversion from metallicity spreads to index distributions via the IMRs. 
Three different ages are selected that show little (5 Gyr) and significant (12 and 13 Gyr) inflection on the IMRs.
We assume the underlying metallicity distribution functions (MDFs) of {10$^6$} model GCs to be box-shaped {and perform the Monte Carlo simulations for index distributions.} 
The simple MDFs should allow us to see the pure effect of the IMR projection.
The indices of each model GC are calculated from its [Fe/H] value via the corresponding IMRs and then typical errors estimated from observation of \citet{Beas04} are randomly added. 
Our simulations with the non-inflected IMRs do not produce bimodal index distributions. 
For example, Fe4531 and Fe5782 of the 5 Gyr models are trapezoid-shaped.
In many cases, the index histograms have a sharp peak with a long tail as the IMRs are broken roughly into parts---the metal-poor, steep section and the metal-rich, shallow section.    
On the other hand, the index distributions produced by the highly inflected IMRs clearly show bimodality. 
For example, the IMRs for G4300, H$\beta$, H$\gamma_{A,F}$, and H$\delta_{A,F}$ of the old GCs have a very shallow slope at [Fe/H] $\simeq$ $-1.0$ between two steeper slopes. 
The inflection brings about a dip on the index distributions by projecting equidistant metallicity intervals onto larger index intervals. 

\begin{figure*}
\includegraphics[angle=0,scale=0.72]{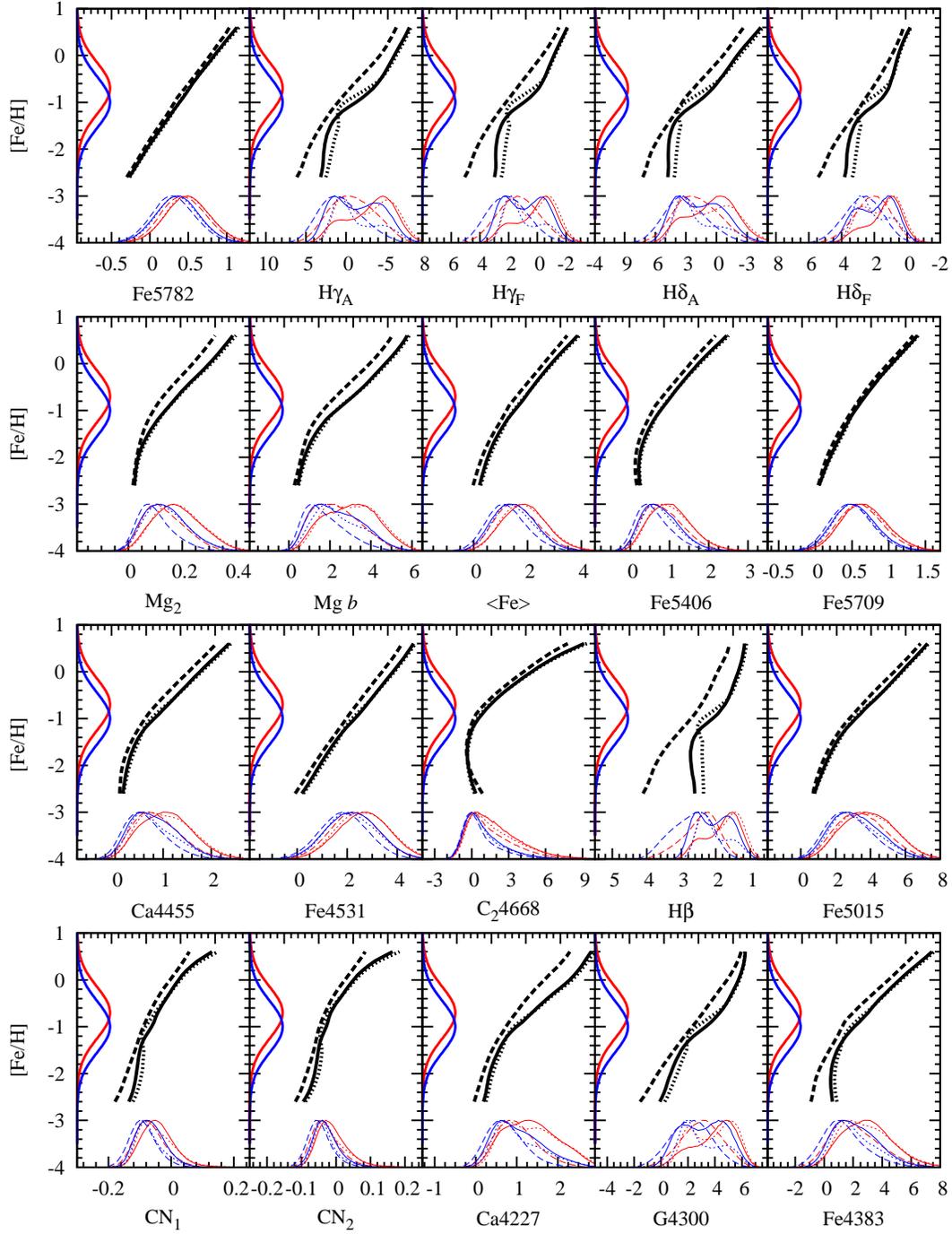}
\caption[]
{
Same as Figure~\ref{1.20} but for the projection tests with the single Gaussian MDFs of $\left<{\rm [Fe/H]}\right>$ = $-0.7$ (red) and $-1.0$ (blue) on various Lick absorption indices. 
{The 10$^6$ model GCs with single Gaussian MDFs are projected on various absorption indices via IMRs of the YEPS model.}
}
\label{1.21}
\end{figure*}

Figure~\ref{1.21} repeats a similar experiment but uses Gaussian MDFs. 
We test for the two Gaussian MDFs of $\langle$[Fe/H]$\rangle$ = $-1.0$ and $-0.7$ with the same dispersion of $\sigma_{[Fe/H]}$ = 0.55. 
In this simulation, we also have applied 10$^6$ model GCs for the given Gaussian MDFs.
Even with Gaussian MDFs, the projected index distributions are double-peaked for G4300, H$\beta$, H$\gamma_{A,F}$, and H$\delta_{A,F}$ of 12-Gyr model GCs.
When a MDF with $\langle$[Fe/H]$\rangle$ = $-1.0$ is used, for example, the KMM algorithm \citep{Ashm94} strongly prefer two peaks for these histograms with p-value $\simeq$ 0.0. 
For the projected H$\beta$ distribution under the assumption of 12~Gyr model, the two peaks are located at 1.681\,{\small \AA} and 2.535\,{\small \AA} with a number fraction of 48.6\,\% and 51.4\,\%, respectively. 
  
It has been claimed that {the bimodal distributions of metal-line indices (e.g., Mg\,$b$ and $[{\rm MgFe}]'$) of} GCs in early-type galaxies are the evidence that GCs have two subsystems with different metallicities \citep{Cohe98, 2003ApJ...592..866C, 2007AJ....133.2015S, 2010ApJ...708.1335W}. 
Interestingly, however, even with unimodal MDF, the index distribution of Mg\,$b$, frequently used as a metallicity indicator, also shows a bimodal index distribution that is consistent with the observations.
Our KMM test for the projected Mg\,$b$ distribution of the 12- and 13-Gyr models supports bimodal distributions with the p-value of 0.0. 
The test implies that the HBs exert an appreciable effect 
not only on the Balmer lines but also on the {\it metallicity-sensitive} lines.
Without assuming two subpopulations in GCs, the projection effect can reproduce the observed bimodal Mg\,$b$ histograms.

\subsubsection{Absorption Index Distributions of M31 Globular Clusters}

 In order to compare the observational index distributions with the simulated ones, we choose M31, the nearest large galaxy, which has spectroscopic surveys of a great number of GCs with reasonably small observation errors.
M31 GCs \citep{Beas04} are of 18 and 33\,\% smaller errors respectively in H$\beta$ and $\langle$Fe$\rangle$ at given absolute magnitudes compared to the M87 GC spectroscopy with Subaru (S. Kim et al. 2012 in preparation).
We excluded young cluster candidates in \citet{Beas04} to avoid young cluster contaminations in the MDF (see blue triangles in Figure~\ref{1.8} to \ref{1.12}). 

\begin{figure*}
\includegraphics[angle=-90,scale=0.6]{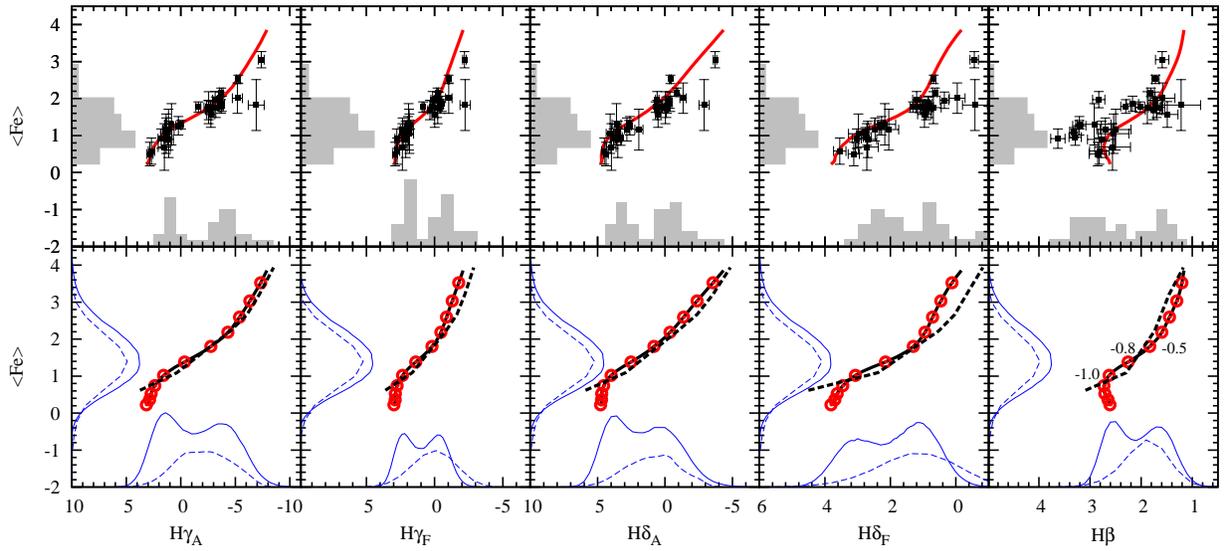}
\caption[]
{Comparison between index distributions of M31 GCs and simulated index distributions. Upper panels are observed histograms
of M31 GCs in H$\beta$, H$\gamma_{A,F}$, H$\delta_{A,F}$, and $\left<\rm Fe \right>$ 
{and the YEPS models (red lines) for each index are overlaid with GCs in M31.}
Lower panels are simulated index distributions with the YEPS and TMB03 models.
Black solid and dashed lines represent the YEPS and TMB03 model {for [$\alpha$/Fe]=0.15}, respectively. 
Blue solid and dashed lines are simulated index distributions based on the YEPS and TMB03 model, respectively. 
Red circles on the solid lines indicate metallicities from [Fe/H] = $-2.6$ to $-0.5$ with a fixed interval $\Delta{\rm [Fe/H]}$ = 0.3.
Three values of ${\rm [Fe/H]}=-1.0$, $-0.8$, and $-0.5$ are indicated in the YEPS model for H$\beta$.
}
\label{1.22}
\end{figure*}

Figure~\ref{1.22} shows the index--index diagrams for $\langle$Fe$\rangle$ and Balmer indices. 
The observed index distributions of M31 GCs and the projection simulations based on the YEPS and TMB03 IMRs are shown as the histograms along the $y$- and $x$-axes, respectively. 
We, again, make a simple assumption of Gaussian MDFs with $\langle$[Fe/H]$\rangle$ = $-1.0$ and $\sigma_{\rm [Fe/H]}$ = 0.55.
As explained in detail in Section 4.2.1, the principle of the MDF projection with IMRs is to transfer SSPs from the metallicity space to the index space. 
Red circles with an uniform metallicity intervals ($\Delta$[Fe/H] $\simeq$ 0.3) on the YEPS IMRs in the bottom panel show this principle of projection.
All Balmer indices show large index space between [Fe/H] = $-1.1$ to $-0.5$ and the corresponding index distributions also show relatively low number frequencies.
However, the $\langle$Fe$\rangle$ index, which keeps relatively constant index intervals at fixed [Fe/H] intervals, does not show a significant change in the projected distribution.
As a result, the projection simulation with the YEPS model reproduce the unimodal $\langle$Fe$\rangle$ and bimodal Balmer line distributions at the same time using a single Gaussian MDF.
We stress that if the observational errors are small enough then the $\langle$Fe$\rangle$ distributions can also show a weak bimodal feature. Indeed, the recent observation of GCs in M31 by \citet{2011AJ....141...61C} show a weak bimodality in the distribution of $\langle$Fe$\rangle$
\citep{skim12}.
The TMB03 model, on the other hand, gives almost straight IMRs and does not reproduce the observed distributions of the $\langle$Fe$\rangle$ and Balmer indices, simultaneously.
No matter how small the errors are, the single Gaussian MDF projection based on the models without systematic HB effect can not reproduce the observed distributions for both $\langle$Fe$\rangle$ and Balmer indices. 

\begin{figure*}
\includegraphics[angle=-90,scale=0.6]{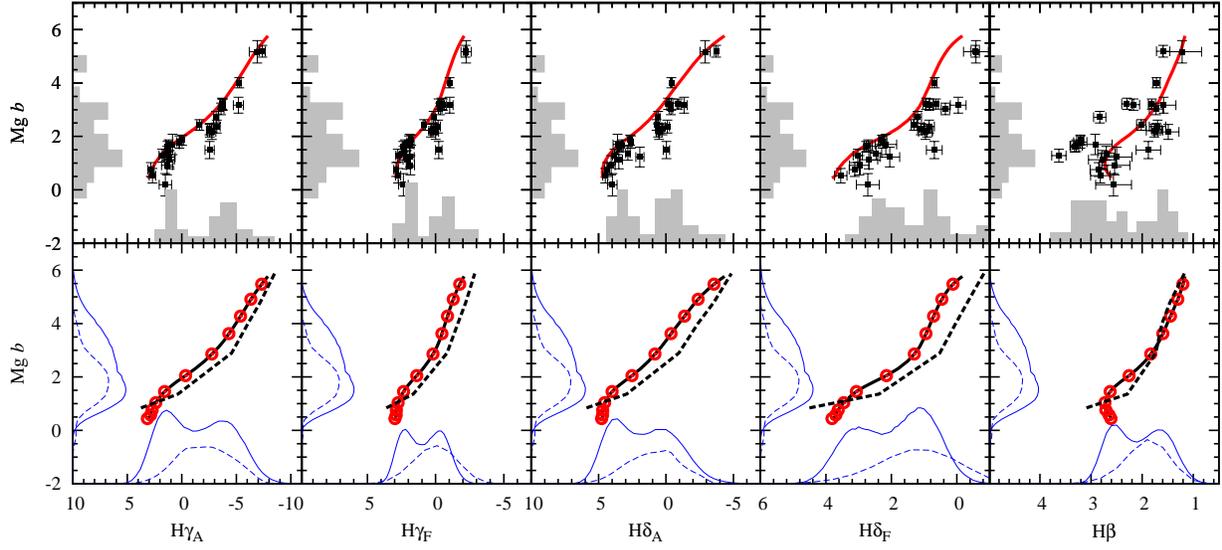}
\caption[]{Same as Figure~\ref{1.22} but for Mg\,$b$ distributions.} 
\label{1.23}
\end{figure*}

Figure~\ref{1.23} shows the index--index diagrams for Mg\,$b$ and Balmer indices. 
The observed and simulated Mg\,$b$ and Balmer distributions are displayed along the $y$- and $x$-axes, respectively.
As discussed in Section 4.2.1, Mg\,$b$ is fairly affected by blue HBs
and thus the distribution of Mg\,$b$ shows weak bimodality.
Many other observations of GCs in early type galaxies also show similar results and the bimodal Mg\,$b$ distributions are often interpreted as the evidence of the bimodal GC MDFs \citep{Cohe98, 2003ApJ...592..866C, 2007AJ....133.2015S, 2010ApJ...708.1335W}. 
The comparison of the YEPS model to M31 GCs demonstrates that the HB effect on Mg\,$b$ is certainly non-negligible. 
Greater caution, therefore, is required in deriving GC metallicity directly from Mg\,$b$.

The spectroscopic line indices, as more detailed probes of stellar populations than broadband colors, 
contain abundant information on the structure of GC systems. 
Our simulations targeting at the old GCs in M31 suggest that two distinct groups found in Balmer and Mg\,$b$ indices
can be due simply to the inflected, nonlinear IMRs.
The result favors a unimodal [Fe/H] distribution of M31 old GCs, in line with the GC structure of extragalactic GC systems suggested by \citet{Yoon06,2011ApJ...743..149Y, 2011ApJ...743..150Y}.
Absorption indices are generally subject to larger observational uncertainties and spectroscopic samples are still small compared to photometric samples with broadband colors. 
More precise spectroscopic observations of greater number of extragalactic GCs by next-generation telescope projects, such as the Giant Magellan Telescope (GMT), Thirty Meter Telescope (TMT) and European Extremely Large Telescope (E-ELT), are highly anticipated in this regard. 

\section{SUMMARY}

We have presented an updated and blown-up version of the Yonsei Evolutionary Populations Synthesis (YEPS) model 
for spectroscopic absorption indices of simple stellar populations.
The characteristics of the YEPS and its applications are summarized as follows.

1. The YEPS has included detailed HB models, which reproduce the observed HB morphology of the Milky Way GCs and varies with respect to metallicity, age, and $\alpha$-elements enhancement.

2. The YEPS has incorporated the $\alpha$-element variation
by using the Y$^2$ stellar library with enhanced $\alpha$-elements and
the response functions of $\alpha$-elements by K05.

3. The YEPS is in good agreement with other EPS models,
except for the indices sensitive to hot stars (CN$_{1}$, CN$_{2}$,
G4300, H$\beta$, H$\gamma$, and H$\delta$).

4. The YEPS reproduces well the observed absorption features of GCs in the MW and M31,
including the strengthened Balmer absorptions by the effect of blue HBs.

{5. When the observed variation of HB morphology with metallicity in the MWGCs is included in the models, Balmer indices of SSPs do not monotonically decrease with increasing metallicity at given age because of blue HB stars in the metal-poor regime. Therefore, the age dating of old stellar systems based on Balmer indices suffers from age degeneracy in the metal-poor regime.}

6. The contribution of HBs to absorption strengths is not limited to
Balmer indices but influences the indices pertinent to iron and
$\alpha$-elements. As a consequence, the most index--metallicity relations of YEPS are inflected
and nonlinear.

7. We have simulated, for the first time, the index distributions of
GCs using the YEPS index--metallicity relations.
The distributions of Balmer and Mg\,$b$ indices constructed under the
unimodal MDF assumption
show clear bimodality, which can be viewed as {a} close analogy to the
well-known bimodality
found in the broadband color distributions of extragalactic GC systems.

\acknowledgments

{We thank the referee for a number of helpful suggestions.}
SJY and YWL acknowledge support from the National Research Foundation of Korea to the Center for Galaxy Evolution Research.
SJY also acknowledges support from Basic Science Research Program (No. 2011-0027247)
through the National Research Foundation (NRF) of Korea grant
funded by the Ministry of Education, Science and Technology (MEST),
and support by the Korea Astronomy and Space Science Institute (KASI) Research Fund 2011 and 2012.
SJY would like to thank Daniel Fabricant, Charles Alcock, Jay Strader, Nelson Caldwell, Dong-Woo Kim, Jae-Sub Hong for their 
hospitality during his stay at Harvard-Smithsonian Center for Astrophysics in 2011--2012.
This work was partially supported by the KASI-Yonsei Joint Research Program (2011 - 2012) for the Frontiers of Astronomy and Space Science.

\clearpage

\clearpage
\clearpage

\clearpage

\clearpage
\clearpage

\clearpage
\clearpage

\clearpage

\clearpage
\clearpage

\clearpage

\clearpage

\clearpage
\clearpage

\clearpage

\clearpage

\clearpage

\clearpage

\clearpage

\clearpage

\clearpage

\clearpage

\clearpage

\clearpage

\clearpage

\clearpage
\clearpage

\clearpage
\clearpage

\clearpage

\clearpage

\clearpage

\begin{deluxetable}{lcc}
\tabletypesize{\scriptsize}
\tablewidth{0pt}
\tablecaption{\label{tab.1.1} Model Input Parameters}
\tablehead{
\colhead{Input Ingredients \& Parameters} &\colhead{Standard Model} &\colhead{Comparison Model}}
\startdata
Stellar Library&Y$^2$ stellar libraries&BaSTI\\
&(\citealt{Kim02}; Lee et al. 2012 {\it in prep.})&\citep{Piet04}\\
Spectral Library & BaSel 3.1 \citep{West02}& BaSel 3.1 \citep{West02}\\
Empirical Fitting Functions&\citet{Wort94}&\citet{Wort94}\\
&\citet{Wort97}&\citet{Wort97}\\
&\citet{Schi07}&\citet{Schi07}\\
Response Functions of $\alpha$-elements&\citet{Korn05}&\citet{Korn05}\\
Initial Mass Function&Salpeter ($x=1.35$)&Salpeter ($x=1.35$)\\
$\alpha$-elements enhancement, [$\alpha$/{\rm Fe}]& 0.0, 0.3, 0.6 & 0.0, 0.4\\
HB mass dispersion, $\sigma_{M} (M_{\odot})$&0.015&0.015\\
\citet{Reim77}'s mass-loss efficiency parameter, $\eta$&0.63&0.40\footnote{For our models with BaSTI stellar library, we have applied the additional mass-loss of 0.01$M_{\odot}$ in order to reproduce HB morphologies of inner halo GCs in the MW at the age of 12~Gyr.}\\
Assumption of the age of inner-halo MWGCs&12 Gyr&12 Gyr\\
\enddata
\end{deluxetable}

\clearpage
\begin{table}
\begin{center}
\caption{\label{tab.1.2}The $\alpha$-element enhanced patterns
of stellar libraries}
\begin{tabular}{cccc}
\tableline\tableline
{$\alpha$-element} & Grevesse &{$Y^2$-Isochrones}& {BaSTI}\\
& {[$\alpha$/{\rm Fe}] = 0.0} &{[$\alpha$/{\rm Fe}] = 0.3}&{[$\alpha$/{\rm Fe}] = 0.4}\\
\tableline
C  & 8.55 & 8.55 & 8.55 \\
N  & 7.97 & 7.97 & 7.97 \\
O  & 8.87 & 9.17 & 9.37 \\
Ne & 8.08 & 8.38 & 8.37 \\
Na & 6.33 & 6.63 & 6.33 \\
Mg & 7.58 & 7.88 & 7.98 \\
Al & 6.47 & 6.17 & 6.47 \\
Si & 7.55 & 7.85 & 7.85 \\
P  & 5.45 & 5.75 & 5.45 \\
S  & 7.21 & 7.51 & 7.54 \\
Cl & 5.50 & 5.80 & 5.50 \\
Ar & 6.52 & 6.82 & 6.52 \\
K  & 5.12 & 5.12 & 5.12 \\
Ca & 6.36 & 6.66 & 6.86 \\
Ti & 5.02 & 5.32 & 5.65 \\
Cr & 5.67 & 5.67 & 5.68 \\
Mn & 5.39 & 5.24 & 5.39 \\
Fe & 7.50 & 7.50 & 7.50 \\
Ni & 6.25 & 6.25 & 6.29 \\
\tableline
\end{tabular}
\end{center}
\tablecomments{The abundance of elements is listed in logarithmic scale $\log$ $N_{el}/N_{H}$ + 12.}

\end{table}

\clearpage

\begin{table}
\begin{center}
\caption{\label{tab.1.3}Minimum Lick indices based on W94 fitting functions}
\begin{tabular}{ccc}
\tableline\tableline
{Lick Index} & {$Y^2$ stellar library} & {BaSTI} \\
\tableline
CN$_{1}$    &  --0.329 &  --0.301 \\*
CN$_{2}$    &  --0.309 &  --0.264 \\*
Ca4227           &  --0.654 &  --0.642 \\*
G4300            &  --5.712 &  --5.712 \\*
Fe4383           &  --4.371 &  --4.377 \\*
Ca4455           &  --0.400 &  --0.404 \\*
Fe4531           &  --1.448 &  --1.417 \\*
Fe4668           &  --6.238 &  --2.156 \\*
H$\beta$     &  --1.726 &  --1.726 \\*
Fe5015           &  --0.838 &  --0.890 \\*
Mg$_{1}$    &  --0.170 &  --0.170 \\*
Mg$_{2}$    &  --0.082 &  --0.082 \\*
Mg\,$b$        &  --1.442 &  --1.444 \\*
Fe5270           &  --2.375 &  --2.350 \\*
Fe5335           &  --0.275 &  --0.294 \\*
Fe5406           &  --0.952 &  --0.951 \\*
Fe5709           &  --1.976 &  --1.976 \\*
Fe5782           &  --0.805 &  --0.804 \\*
NaD             &   0.000 &   0.000 \\*
TiO$_{1}$         &  --0.068 &  --0.018 \\*
TiO$_{2}$         &  --0.055 &  --0.008 \\*
H$\gamma_{A}$   & --12.269 & --12.152 \\*
H$\gamma_{F}$   &  --4.292 &  --4.278 \\*
H$\delta_{A}$   &  --9.384 &  --9.139 \\*
H$\delta_{F}$   &  --2.359 &  --2.249 \\*
\tableline
\end{tabular}
\end{center}
\end{table}

\clearpage

\begin{deluxetable}{lccc}
\tabletypesize{\scriptsize}
\tablewidth{0pt}
\tablecaption{\label{tab.1.4}Input Parameters of CMD model}
\tablehead{
\colhead{Parameters} &\colhead{47~Tuc} &\colhead{NGC~6284} &\colhead{M67}}
\startdata
Initial Mass Function&Salpeter ($x=1.35$)&Salpeter ($x=1.35$) & Salpeter ($x=1.35$)\\
$\alpha$-elements enhancement, [$\alpha$/{\rm Fe}]& 0.3& 0.3 & 0.3\\
HB mass dispersion, $\sigma_{M} (M_{\odot})$&0.015&0.015&0.015\\
\citet{Reim77}'s mass-loss efficiency parameter, $\eta$&0.63&0.63&0.63\\
Distance modulus, ($V-M_V$) (mag)&13.35&16.70&9.95\\
Galactic reddening, $E(B-V)$ (mag)&0.01&0.32&0.04\\
Metal abundance, [Fe/H]&$-$0.73&$-$1.50&0.00\\
Absolute age, $t$ (Gyr)&11.9&13.1&3.5\\
\enddata
\end{deluxetable}

\clearpage

\begin{deluxetable}{ccccccccccccc}
\tabletypesize{\scriptsize}
\tablewidth{0pt}
\tablecaption{\label{tab.1.5}LICK absorption indices of YEPS simple stellar population model for [$\alpha$/Fe]=0.0 (fitting functions of W94 and W97).}
\tablehead{
\colhead{Age = 12.0} & \colhead{} & \colhead{} & \colhead{} & \colhead{} & \colhead{} & \colhead{} & \colhead{} & \colhead{} & \colhead{} & \colhead{} & \colhead{} & \colhead{}}
\startdata
 [Fe/H] & CN$_1$ & CN$_2$ & Ca4227 & G4300 & Fe4383 & Ca4455 & Fe4531 & C$_2$4668 & H$\beta$ & Fe5015 & Mg$_1$ & Mg$_2$ \\ 
\hline 
 -2.50 &  -0.132 &  -0.089 &   0.213 &   0.033 &   0.451 &   0.166 &   0.258 &   0.156 &   2.672 &   0.798 &   0.007 &   0.021  \\ 
 -2.40 &  -0.126 &  -0.082 &   0.230 &   0.207 &   0.426 &   0.181 &   0.378 &   0.010 &   2.673 &   0.885 &   0.007 &   0.023  \\ 
 -2.30 &  -0.121 &  -0.076 &   0.242 &   0.342 &   0.401 &   0.196 &   0.490 &  -0.100 &   2.693 &   0.982 &   0.006 &   0.024  \\ 
 -2.20 &  -0.118 &  -0.072 &   0.256 &   0.465 &   0.393 &   0.214 &   0.600 &  -0.202 &   2.719 &   1.086 &   0.006 &   0.026  \\ 
 -2.10 &  -0.116 &  -0.069 &   0.265 &   0.563 &   0.390 &   0.232 &   0.703 &  -0.285 &   2.758 &   1.188 &   0.006 &   0.028  \\ 
 -2.00 &  -0.114 &  -0.065 &   0.281 &   0.688 &   0.421 &   0.254 &   0.813 &  -0.359 &   2.775 &   1.292 &   0.007 &   0.032  \\ 
 -1.90 &  -0.112 &  -0.063 &   0.298 &   0.810 &   0.468 &   0.278 &   0.922 &  -0.417 &   2.794 &   1.399 &   0.008 &   0.035  \\ 
 -1.80 &  -0.110 &  -0.060 &   0.320 &   0.958 &   0.543 &   0.308 &   1.035 &  -0.454 &   2.792 &   1.514 &   0.010 &   0.040  \\ 
 -1.70 &  -0.107 &  -0.058 &   0.344 &   1.104 &   0.637 &   0.341 &   1.149 &  -0.465 &   2.793 &   1.633 &   0.012 &   0.046  \\ 
 -1.60 &  -0.104 &  -0.055 &   0.378 &   1.301 &   0.768 &   0.382 &   1.273 &  -0.453 &   2.765 &   1.771 &   0.015 &   0.053  \\ 
 -1.50 &  -0.099 &  -0.052 &   0.418 &   1.538 &   0.931 &   0.431 &   1.405 &  -0.406 &   2.712 &   1.926 &   0.018 &   0.061  \\ 
 -1.40 &  -0.094 &  -0.049 &   0.467 &   1.813 &   1.119 &   0.489 &   1.545 &  -0.312 &   2.630 &   2.103 &   0.021 &   0.070  \\ 
 -1.30 &  -0.089 &  -0.045 &   0.517 &   2.080 &   1.320 &   0.550 &   1.683 &  -0.179 &   2.556 &   2.292 &   0.026 &   0.081  \\ 
 -1.20 &  -0.084 &  -0.042 &   0.566 &   2.339 &   1.527 &   0.613 &   1.816 &  -0.018 &   2.499 &   2.487 &   0.030 &   0.092  \\ 
 -1.10 &  -0.076 &  -0.037 &   0.634 &   2.718 &   1.799 &   0.695 &   1.976 &   0.202 &   2.376 &   2.719 &   0.035 &   0.104  \\ 
 -1.00 &  -0.066 &  -0.030 &   0.705 &   3.147 &   2.094 &   0.785 &   2.140 &   0.468 &   2.226 &   2.962 &   0.040 &   0.117  \\ 
 -0.90 &  -0.055 &  -0.022 &   0.787 &   3.606 &   2.448 &   0.886 &   2.313 &   0.799 &   2.068 &   3.227 &   0.046 &   0.131  \\ 
 -0.80 &  -0.045 &  -0.015 &   0.868 &   4.028 &   2.844 &   0.989 &   2.479 &   1.176 &   1.939 &   3.492 &   0.053 &   0.145  \\ 
 -0.70 &  -0.035 &  -0.007 &   0.959 &   4.417 &   3.277 &   1.099 &   2.648 &   1.613 &   1.815 &   3.773 &   0.060 &   0.158  \\ 
 -0.60 &  -0.028 &  -0.001 &   1.040 &   4.695 &   3.658 &   1.200 &   2.799 &   2.054 &   1.730 &   4.038 &   0.068 &   0.172  \\ 
 -0.50 &  -0.018 &   0.008 &   1.108 &   4.946 &   3.952 &   1.295 &   2.932 &   2.463 &   1.660 &   4.332 &   0.074 &   0.185  \\ 
 -0.40 &  -0.003 &   0.023 &   1.182 &   5.171 &   4.269 &   1.394 &   3.069 &   2.940 &   1.604 &   4.634 &   0.079 &   0.199  \\ 
 -0.30 &   0.009 &   0.036 &   1.286 &   5.346 &   4.696 &   1.499 &   3.223 &   3.516 &   1.556 &   4.886 &   0.086 &   0.217  \\ 
 -0.20 &   0.020 &   0.048 &   1.392 &   5.509 &   5.117 &   1.604 &   3.374 &   4.078 &   1.501 &   5.125 &   0.096 &   0.234  \\ 
 -0.10 &   0.031 &   0.060 &   1.500 &   5.659 &   5.530 &   1.708 &   3.521 &   4.646 &   1.446 &   5.355 &   0.105 &   0.252  \\ 
  0.00 &   0.043 &   0.072 &   1.607 &   5.798 &   5.943 &   1.813 &   3.666 &   5.243 &   1.392 &   5.581 &   0.116 &   0.269  \\ 
  0.10 &   0.056 &   0.086 &   1.712 &   5.924 &   6.353 &   1.917 &   3.809 &   5.875 &   1.344 &   5.805 &   0.126 &   0.287  \\ 
  0.20 &   0.071 &   0.102 &   1.816 &   6.037 &   6.767 &   2.023 &   3.953 &   6.565 &   1.302 &   6.034 &   0.136 &   0.303  \\ 
  0.30 &   0.087 &   0.119 &   1.918 &   6.132 &   7.177 &   2.128 &   4.095 &   7.308 &   1.267 &   6.266 &   0.146 &   0.320  \\ 
  0.40 &   0.108 &   0.141 &   2.018 &   6.209 &   7.590 &   2.236 &   4.241 &   8.136 &   1.238 &   6.516 &   0.155 &   0.336  \\ 
  0.50 &   0.134 &   0.169 &   2.113 &   6.260 &   8.007 &   2.344 &   4.389 &   9.057 &   1.213 &   6.779 &   0.164 &   0.352  \\ 
\hline
 [Fe/H] & Mg\,$b$ & Fe5270 & Fe5335 & Fe5406 & Fe5709 & Fe5782 & NaD & H$\gamma_A$ & H$\gamma_F$ & H$\delta_A$ & H$\delta_F$ & \\
\hline
 -2.50 &   0.411 &  -0.137 &   0.606 &   0.227 &   0.066 &  -0.238 &   1.001 &   3.106 &   2.968 &   4.869 &   3.830 & \\
 -2.40 &   0.446 &  -0.031 &   0.592 &   0.216 &   0.089 &  -0.201 &   0.976 &   2.966 &   2.911 &   4.807 &   3.764 & \\
 -2.30 &   0.487 &   0.070 &   0.584 &   0.211 &   0.113 &  -0.163 &   0.961 &   2.881 &   2.896 &   4.782 &   3.739 & \\
 -2.20 &   0.531 &   0.172 &   0.582 &   0.211 &   0.137 &  -0.126 &   0.952 &   2.837 &   2.898 &   4.779 &   3.721 & \\
 -2.10 &   0.578 &   0.271 &   0.585 &   0.216 &   0.163 &  -0.088 &   0.949 &   2.838 &   2.927 &   4.801 &   3.721 & \\
 -2.00 &   0.634 &   0.374 &   0.597 &   0.227 &   0.191 &  -0.050 &   0.952 &   2.784 &   2.915 &   4.771 &   3.681 & \\
 -1.90 &   0.696 &   0.478 &   0.617 &   0.245 &   0.220 &  -0.010 &   0.963 &   2.729 &   2.901 &   4.738 &   3.638 & \\
 -1.80 &   0.771 &   0.586 &   0.648 &   0.271 &   0.252 &   0.031 &   0.984 &   2.610 &   2.847 &   4.657 &   3.565 & \\
 -1.70 &   0.855 &   0.696 &   0.689 &   0.305 &   0.284 &   0.073 &   1.017 &   2.481 &   2.789 &   4.564 &   3.488 & \\
 -1.60 &   0.951 &   0.814 &   0.742 &   0.348 &   0.317 &   0.115 &   1.065 &   2.247 &   2.670 &   4.383 &   3.356 & \\
 -1.50 &   1.062 &   0.938 &   0.808 &   0.399 &   0.351 &   0.159 &   1.127 &   1.909 &   2.495 &   4.119 &   3.178 & \\
 -1.40 &   1.196 &   1.073 &   0.890 &   0.461 &   0.388 &   0.205 &   1.208 &   1.484 &   2.271 &   3.795 &   2.969 & \\
 -1.30 &   1.348 &   1.208 &   0.980 &   0.528 &   0.425 &   0.250 &   1.303 &   1.046 &   2.052 &   3.479 &   2.779 & \\
 -1.20 &   1.498 &   1.338 &   1.075 &   0.597 &   0.460 &   0.293 &   1.410 &   0.637 &   1.861 &   3.160 &   2.591 & \\
 -1.10 &   1.678 &   1.487 &   1.189 &   0.679 &   0.498 &   0.338 &   1.533 &  -0.051 &   1.521 &   2.619 &   2.280 & \\
 -1.00 &   1.878 &   1.640 &   1.314 &   0.765 &   0.538 &   0.382 &   1.665 &  -0.868 &   1.108 &   1.979 &   1.911 & \\
 -0.90 &   2.113 &   1.804 &   1.456 &   0.863 &   0.581 &   0.428 &   1.815 &  -1.796 &   0.650 &   1.296 &   1.546 & \\
 -0.80 &   2.352 &   1.968 &   1.603 &   0.964 &   0.625 &   0.474 &   1.976 &  -2.687 &   0.225 &   0.650 &   1.248 & \\
 -0.70 &   2.597 &   2.138 &   1.764 &   1.075 &   0.675 &   0.521 &   2.153 &  -3.548 &  -0.167 &  -0.022 &   0.996 & \\
 -0.60 &   2.825 &   2.294 &   1.918 &   1.182 &   0.722 &   0.565 &   2.332 &  -4.178 &  -0.423 &  -0.563 &   0.834 & \\
 -0.50 &   3.042 &   2.429 &   2.047 &   1.279 &   0.772 &   0.605 &   2.465 &  -4.757 &  -0.631 &  -1.035 &   0.701 & \\
 -0.40 &   3.251 &   2.568 &   2.183 &   1.383 &   0.821 &   0.652 &   2.607 &  -5.301 &  -0.815 &  -1.513 &   0.573 & \\
 -0.30 &   3.458 &   2.723 &   2.358 &   1.501 &   0.867 &   0.705 &   2.828 &  -5.747 &  -0.965 &  -1.984 &   0.468 & \\
 -0.20 &   3.658 &   2.871 &   2.534 &   1.619 &   0.919 &   0.756 &   3.057 &  -6.181 &  -1.109 &  -2.441 &   0.364 & \\
 -0.10 &   3.857 &   3.016 &   2.712 &   1.737 &   0.974 &   0.806 &   3.296 &  -6.598 &  -1.251 &  -2.886 &   0.259 & \\
  0.00 &   4.064 &   3.158 &   2.893 &   1.857 &   1.032 &   0.855 &   3.542 &  -7.004 &  -1.392 &  -3.331 &   0.152 & \\
  0.10 &   4.272 &   3.297 &   3.078 &   1.979 &   1.089 &   0.903 &   3.804 &  -7.392 &  -1.536 &  -3.771 &   0.041 & \\
  0.20 &   4.473 &   3.437 &   3.270 &   2.105 &   1.150 &   0.952 &   4.078 &  -7.769 &  -1.688 &  -4.212 &  -0.077 & \\
  0.30 &   4.672 &   3.575 &   3.468 &   2.234 &   1.210 &   0.999 &   4.368 &  -8.122 &  -1.841 &  -4.635 &  -0.197 & \\
  0.40 &   4.864 &   3.716 &   3.675 &   2.367 &   1.271 &   1.051 &   4.672 &  -8.468 &  -2.004 &  -5.057 &  -0.331 & \\
  0.50 &   5.044 &   3.864 &   3.890 &   2.500 &   1.334 &   1.105 &   4.996 &  -8.791 &  -2.165 &  -5.469 &  -0.481 & \\
\enddata
\tablecomments{The entire data of Table~\ref{tab.1.5} are available at http://web.yonsei.ac.kr/cosmic/data/YEPS.htm.}
\end{deluxetable}

\clearpage

\begin{deluxetable}{ccccccccccccc}
\tabletypesize{\scriptsize}
\tablewidth{0pt}
\tablecaption{\label{tab.1.6}LICK absorption indices of YEPS simple stellar population model for [$\alpha$/Fe]=0.3 (fitting functions of W94 and W97).}
\tablehead{
\colhead{Age = 12.0} & \colhead{} & \colhead{} & \colhead{} & \colhead{} & \colhead{} & \colhead{} & \colhead{} & \colhead{} & \colhead{} & \colhead{} & \colhead{} & \colhead{}}
\startdata
 [Fe/H] & CN$_1$ & CN$_2$ & Ca4227 & G4300 & Fe4383 & Ca4455 & Fe4531 & C$_2$4668 & H$\beta$ & Fe5015 & Mg$_1$ & Mg$_2$ \\ 
\hline 
 -2.50 &  -0.125 &  -0.084 &   0.235 &   0.122 &   0.611 &   0.184 &   0.371 &   0.242 &   2.561 &   0.930 &   0.008 &   0.027   \\ 
 -2.40 &  -0.120 &  -0.077 &   0.274 &   0.264 &   0.570 &   0.199 &   0.500 &   0.116 &   2.576 &   1.033 &   0.008 &   0.029   \\ 
 -2.30 &  -0.117 &  -0.071 &   0.298 &   0.372 &   0.529 &   0.215 &   0.618 &   0.010 &   2.601 &   1.140 &   0.008 &   0.031   \\ 
 -2.20 &  -0.114 &  -0.067 &   0.322 &   0.475 &   0.502 &   0.231 &   0.733 &  -0.080 &   2.626 &   1.251 &   0.008 &   0.033   \\ 
 -2.10 &  -0.110 &  -0.062 &   0.356 &   0.606 &   0.505 &   0.252 &   0.855 &  -0.157 &   2.634 &   1.367 &   0.009 &   0.037   \\ 
 -2.00 &  -0.108 &  -0.059 &   0.399 &   0.738 &   0.517 &   0.275 &   0.977 &  -0.214 &   2.644 &   1.484 &   0.011 &   0.042   \\ 
 -1.90 &  -0.106 &  -0.056 &   0.450 &   0.872 &   0.549 &   0.301 &   1.101 &  -0.248 &   2.654 &   1.607 &   0.014 &   0.049   \\ 
 -1.80 &  -0.104 &  -0.054 &   0.507 &   1.016 &   0.602 &   0.332 &   1.230 &  -0.265 &   2.662 &   1.738 &   0.017 &   0.057   \\ 
 -1.70 &  -0.103 &  -0.052 &   0.568 &   1.164 &   0.674 &   0.366 &   1.358 &  -0.257 &   2.666 &   1.875 &   0.021 &   0.066   \\ 
 -1.60 &  -0.101 &  -0.050 &   0.640 &   1.345 &   0.777 &   0.407 &   1.493 &  -0.222 &   2.644 &   2.030 &   0.025 &   0.076   \\ 
 -1.50 &  -0.100 &  -0.049 &   0.704 &   1.482 &   0.874 &   0.447 &   1.616 &  -0.163 &   2.646 &   2.185 &   0.029 &   0.087   \\ 
 -1.40 &  -0.097 &  -0.048 &   0.775 &   1.696 &   1.023 &   0.498 &   1.754 &  -0.095 &   2.624 &   2.362 &   0.033 &   0.099   \\ 
 -1.30 &  -0.090 &  -0.044 &   0.863 &   2.032 &   1.230 &   0.564 &   1.911 &   0.019 &   2.526 &   2.569 &   0.037 &   0.112   \\ 
 -1.20 &  -0.079 &  -0.038 &   0.972 &   2.481 &   1.474 &   0.647 &   2.089 &   0.179 &   2.348 &   2.805 &   0.041 &   0.126   \\ 
 -1.10 &  -0.067 &  -0.031 &   1.104 &   2.994 &   1.752 &   0.744 &   2.279 &   0.400 &   2.137 &   3.068 &   0.046 &   0.142   \\ 
 -1.00 &  -0.059 &  -0.026 &   1.275 &   3.464 &   2.074 &   0.847 &   2.470 &   0.659 &   1.956 &   3.348 &   0.053 &   0.158   \\ 
 -0.90 &  -0.055 &  -0.025 &   1.480 &   3.836 &   2.386 &   0.950 &   2.648 &   0.922 &   1.815 &   3.622 &   0.060 &   0.174   \\ 
 -0.80 &  -0.053 &  -0.025 &   1.638 &   4.089 &   2.626 &   1.036 &   2.796 &   1.209 &   1.728 &   3.864 &   0.067 &   0.190   \\ 
 -0.70 &  -0.050 &  -0.023 &   1.786 &   4.327 &   2.793 &   1.117 &   2.932 &   1.489 &   1.650 &   4.129 &   0.074 &   0.206   \\ 
 -0.60 &  -0.041 &  -0.014 &   1.937 &   4.583 &   2.916 &   1.200 &   3.070 &   1.796 &   1.579 &   4.448 &   0.079 &   0.222   \\ 
 -0.50 &  -0.034 &  -0.006 &   2.124 &   4.766 &   3.183 &   1.291 &   3.229 &   2.240 &   1.529 &   4.705 &   0.087 &   0.242   \\ 
 -0.40 &  -0.028 &  -0.001 &   2.299 &   4.942 &   3.432 &   1.375 &   3.377 &   2.642 &   1.480 &   4.940 &   0.095 &   0.262   \\ 
 -0.30 &  -0.022 &   0.005 &   2.466 &   5.121 &   3.687 &   1.461 &   3.526 &   3.066 &   1.436 &   5.182 &   0.104 &   0.280   \\ 
 -0.20 &  -0.014 &   0.012 &   2.647 &   5.315 &   3.918 &   1.544 &   3.682 &   3.513 &   1.391 &   5.443 &   0.113 &   0.298   \\ 
 -0.10 &  -0.006 &   0.020 &   2.815 &   5.503 &   4.157 &   1.628 &   3.838 &   3.984 &   1.348 &   5.709 &   0.121 &   0.315   \\ 
  0.00 &   0.004 &   0.030 &   2.966 &   5.684 &   4.423 &   1.717 &   3.996 &   4.488 &   1.308 &   5.986 &   0.129 &   0.331   \\ 
  0.10 &   0.014 &   0.039 &   3.079 &   5.824 &   4.721 &   1.806 &   4.144 &   4.988 &   1.264 &   6.246 &   0.136 &   0.346   \\ 
  0.20 &   0.027 &   0.051 &   3.169 &   5.951 &   5.055 &   1.902 &   4.290 &   5.529 &   1.226 &   6.519 &   0.142 &   0.359   \\ 
  0.30 &   0.042 &   0.067 &   3.241 &   6.056 &   5.425 &   2.004 &   4.434 &   6.111 &   1.192 &   6.797 &   0.147 &   0.372   \\ 
  0.40 &   0.059 &   0.084 &   3.294 &   6.131 &   5.785 &   2.102 &   4.558 &   6.760 &   1.168 &   7.062 &   0.152 &   0.383   \\ 
  0.50 &   0.079 &   0.105 &   3.349 &   6.172 &   6.135 &   2.199 &   4.667 &   7.486 &   1.150 &   7.301 &   0.159 &   0.396   \\ 
\hline
 [Fe/H] & Mg\,$b$ & Fe5270 & Fe5335 & Fe5406 & Fe5709 & Fe5782 & NaD & H$\gamma_A$ & H$\gamma_F$ & H$\delta_A$ & H$\delta_F$ & \\
\hline
 -2.50 &   0.556 &  -0.050 &   0.649 &   0.237 &   0.080 &  -0.229 &   1.032 &   2.945 &   2.886 &   4.603 &   3.685 & \\
 -2.40 &   0.611 &   0.060 &   0.636 &   0.228 &   0.104 &  -0.189 &   1.033 &   2.870 &   2.870 &   4.604 &   3.656 & \\
 -2.30 &   0.672 &   0.165 &   0.627 &   0.222 &   0.127 &  -0.151 &   1.034 &   2.843 &   2.887 &   4.631 &   3.655 & \\
 -2.20 &   0.737 &   0.269 &   0.622 &   0.221 &   0.152 &  -0.113 &   1.042 &   2.821 &   2.903 &   4.657 &   3.652 & \\
 -2.10 &   0.812 &   0.379 &   0.628 &   0.228 &   0.178 &  -0.074 &   1.060 &   2.743 &   2.879 &   4.642 &   3.614 & \\
 -2.00 &   0.895 &   0.489 &   0.641 &   0.242 &   0.207 &  -0.035 &   1.093 &   2.656 &   2.848 &   4.632 &   3.576 & \\
 -1.90 &   0.992 &   0.602 &   0.665 &   0.265 &   0.237 &   0.006 &   1.142 &   2.553 &   2.808 &   4.616 &   3.536 & \\
 -1.80 &   1.101 &   0.718 &   0.699 &   0.295 &   0.268 &   0.049 &   1.207 &   2.441 &   2.760 &   4.592 &   3.482 & \\
 -1.70 &   1.226 &   0.834 &   0.741 &   0.333 &   0.299 &   0.091 &   1.288 &   2.296 &   2.702 &   4.543 &   3.426 & \\
 -1.60 &   1.376 &   0.959 &   0.797 &   0.379 &   0.332 &   0.134 &   1.386 &   2.070 &   2.598 &   4.431 &   3.337 & \\
 -1.50 &   1.539 &   1.075 &   0.856 &   0.428 &   0.363 &   0.176 &   1.496 &   1.921 &   2.549 &   4.371 &   3.297 & \\
 -1.40 &   1.709 &   1.197 &   0.924 &   0.480 &   0.394 &   0.218 &   1.612 &   1.671 &   2.442 &   4.205 &   3.172 & \\
 -1.30 &   1.903 &   1.331 &   1.004 &   0.540 &   0.425 &   0.259 &   1.739 &   1.134 &   2.182 &   3.777 &   2.898 & \\
 -1.20 &   2.133 &   1.479 &   1.100 &   0.608 &   0.460 &   0.302 &   1.879 &   0.326 &   1.751 &   3.130 &   2.479 & \\
 -1.10 &   2.419 &   1.639 &   1.210 &   0.687 &   0.499 &   0.347 &   2.036 &  -0.636 &   1.239 &   2.418 &   2.024 & \\
 -1.00 &   2.744 &   1.804 &   1.330 &   0.774 &   0.544 &   0.392 &   2.219 &  -1.593 &   0.756 &   1.797 &   1.680 & \\
 -0.90 &   3.069 &   1.960 &   1.449 &   0.862 &   0.592 &   0.437 &   2.416 &  -2.361 &   0.377 &   1.302 &   1.459 & \\
 -0.80 &   3.364 &   2.088 &   1.554 &   0.941 &   0.636 &   0.477 &   2.618 &  -2.823 &   0.168 &   0.980 &   1.355 & \\
 -0.70 &   3.650 &   2.199 &   1.639 &   1.013 &   0.682 &   0.511 &   2.793 &  -3.220 &  -0.005 &   0.733 &   1.287 & \\
 -0.60 &   3.929 &   2.301 &   1.711 &   1.084 &   0.733 &   0.550 &   2.929 &  -3.628 &  -0.172 &   0.510 &   1.229 & \\
 -0.50 &   4.200 &   2.428 &   1.832 &   1.176 &   0.778 &   0.600 &   3.175 &  -3.889 &  -0.284 &   0.287 &   1.194 & \\
 -0.40 &   4.446 &   2.546 &   1.950 &   1.261 &   0.828 &   0.646 &   3.422 &  -4.131 &  -0.399 &   0.085 &   1.147 & \\
 -0.30 &   4.676 &   2.666 &   2.074 &   1.349 &   0.882 &   0.691 &   3.683 &  -4.381 &  -0.513 &  -0.138 &   1.098 & \\
 -0.20 &   4.893 &   2.782 &   2.191 &   1.433 &   0.939 &   0.736 &   3.970 &  -4.593 &  -0.633 &  -0.312 &   1.057 & \\
 -0.10 &   5.114 &   2.899 &   2.318 &   1.519 &   0.997 &   0.780 &   4.276 &  -4.804 &  -0.759 &  -0.498 &   1.012 & \\
  0.00 &   5.333 &   3.024 &   2.457 &   1.614 &   1.057 &   0.825 &   4.593 &  -5.038 &  -0.899 &  -0.722 &   0.954 & \\
  0.10 &   5.547 &   3.150 &   2.612 &   1.716 &   1.113 &   0.871 &   4.915 &  -5.280 &  -1.045 &  -0.981 &   0.882 & \\
  0.20 &   5.747 &   3.284 &   2.781 &   1.826 &   1.171 &   0.919 &   5.239 &  -5.561 &  -1.203 &  -1.304 &   0.794 & \\
  0.30 &   5.929 &   3.426 &   2.965 &   1.946 &   1.231 &   0.971 &   5.566 &  -5.881 &  -1.376 &  -1.691 &   0.685 & \\
  0.40 &   6.089 &   3.563 &   3.147 &   2.064 &   1.285 &   1.019 &   5.876 &  -6.171 &  -1.536 &  -2.071 &   0.571 & \\
  0.50 &   6.214 &   3.698 &   3.326 &   2.179 &   1.336 &   1.067 &   6.168 &  -6.443 &  -1.694 &  -2.448 &   0.439 & \\
\enddata
\tablecomments{The entire data of Table~\ref{tab.1.6} are available at http://web.yonsei.ac.kr/cosmic/data/YEPS.htm.}
\end{deluxetable}

\clearpage

\begin{deluxetable}{ccccccccccccc}
\tabletypesize{\scriptsize}
\tablewidth{0pt}
\tablecaption{\label{tab.1.7}LICK absorption indices of YEPS simple stellar population model for [$\alpha$/Fe]=0.6 (fitting functions of W94 and W97).}
\tablehead{
\colhead{Age = 12.0} & \colhead{} & \colhead{} & \colhead{} & \colhead{} & \colhead{} & \colhead{} & \colhead{} & \colhead{} & \colhead{} & \colhead{} & \colhead{} & \colhead{}}
\startdata
 [Fe/H] & CN$_1$ & CN$_2$ & Ca4227 & G4300 & Fe4383 & Ca4455 & Fe4531 & C$_2$4668 & H$\beta$ & Fe5015 & Mg$_1$ & Mg$_2$ \\ 
\hline 
 -2.50 &  -0.126 &  -0.082 &   0.207 &  -0.015 &   0.733 &   0.206 &   0.464 &   0.409 &   2.587 &   1.073 &   0.010 &   0.031  \\ 
 -2.40 &  -0.123 &  -0.076 &   0.264 &   0.092 &   0.656 &   0.218 &   0.593 &   0.288 &   2.622 &   1.179 &   0.011 &   0.035  \\ 
 -2.30 &  -0.119 &  -0.071 &   0.319 &   0.216 &   0.583 &   0.227 &   0.721 &   0.156 &   2.630 &   1.284 &   0.010 &   0.037  \\ 
 -2.20 &  -0.115 &  -0.066 &   0.378 &   0.340 &   0.519 &   0.239 &   0.848 &   0.043 &   2.635 &   1.393 &   0.011 &   0.041  \\ 
 -2.10 &  -0.113 &  -0.062 &   0.443 &   0.460 &   0.477 &   0.254 &   0.978 &  -0.040 &   2.651 &   1.507 &   0.013 &   0.047  \\ 
 -2.00 &  -0.110 &  -0.059 &   0.518 &   0.600 &   0.473 &   0.275 &   1.114 &  -0.092 &   2.663 &   1.635 &   0.016 &   0.055  \\ 
 -1.90 &  -0.107 &  -0.055 &   0.603 &   0.769 &   0.516 &   0.306 &   1.259 &  -0.102 &   2.657 &   1.780 &   0.020 &   0.065  \\ 
 -1.80 &  -0.104 &  -0.052 &   0.696 &   0.951 &   0.589 &   0.342 &   1.408 &  -0.083 &   2.648 &   1.938 &   0.025 &   0.077  \\ 
 -1.70 &  -0.102 &  -0.049 &   0.787 &   1.119 &   0.678 &   0.381 &   1.553 &  -0.045 &   2.657 &   2.103 &   0.030 &   0.090  \\ 
 -1.60 &  -0.096 &  -0.045 &   0.896 &   1.396 &   0.827 &   0.433 &   1.716 &   0.009 &   2.605 &   2.294 &   0.035 &   0.104  \\ 
 -1.50 &  -0.087 &  -0.040 &   1.025 &   1.782 &   1.015 &   0.497 &   1.896 &   0.077 &   2.481 &   2.504 &   0.040 &   0.119  \\ 
 -1.40 &  -0.077 &  -0.034 &   1.175 &   2.247 &   1.215 &   0.574 &   2.089 &   0.182 &   2.295 &   2.740 &   0.045 &   0.134  \\ 
 -1.30 &  -0.066 &  -0.027 &   1.363 &   2.730 &   1.433 &   0.661 &   2.293 &   0.334 &   2.094 &   3.004 &   0.051 &   0.152  \\ 
 -1.20 &  -0.060 &  -0.024 &   1.631 &   3.099 &   1.637 &   0.750 &   2.501 &   0.484 &   1.926 &   3.297 &   0.057 &   0.169  \\ 
 -1.10 &  -0.060 &  -0.026 &   1.892 &   3.378 &   1.791 &   0.831 &   2.677 &   0.629 &   1.802 &   3.551 &   0.064 &   0.187  \\ 
 -1.00 &  -0.061 &  -0.030 &   2.120 &   3.587 &   1.880 &   0.898 &   2.821 &   0.779 &   1.714 &   3.773 &   0.071 &   0.205  \\ 
 -0.90 &  -0.062 &  -0.032 &   2.347 &   3.814 &   1.887 &   0.961 &   2.956 &   0.913 &   1.621 &   4.019 &   0.077 &   0.223  \\ 
 -0.80 &  -0.057 &  -0.028 &   2.585 &   4.059 &   1.863 &   1.029 &   3.099 &   1.106 &   1.533 &   4.316 &   0.082 &   0.243  \\ 
 -0.70 &  -0.056 &  -0.027 &   2.838 &   4.223 &   1.981 &   1.102 &   3.251 &   1.397 &   1.473 &   4.548 &   0.089 &   0.264  \\ 
 -0.60 &  -0.056 &  -0.028 &   3.088 &   4.383 &   2.098 &   1.175 &   3.400 &   1.672 &   1.413 &   4.770 &   0.099 &   0.287  \\ 
 -0.50 &  -0.056 &  -0.029 &   3.334 &   4.545 &   2.225 &   1.249 &   3.548 &   1.960 &   1.357 &   4.992 &   0.108 &   0.309  \\ 
 -0.40 &  -0.056 &  -0.029 &   3.602 &   4.717 &   2.329 &   1.321 &   3.702 &   2.256 &   1.301 &   5.226 &   0.119 &   0.331  \\ 
 -0.30 &  -0.053 &  -0.027 &   3.857 &   4.923 &   2.452 &   1.398 &   3.866 &   2.594 &   1.257 &   5.497 &   0.128 &   0.352  \\ 
 -0.20 &  -0.048 &  -0.023 &   4.093 &   5.150 &   2.583 &   1.475 &   4.037 &   2.977 &   1.225 &   5.803 &   0.137 &   0.370  \\ 
 -0.10 &  -0.042 &  -0.018 &   4.295 &   5.374 &   2.748 &   1.557 &   4.209 &   3.386 &   1.202 &   6.124 &   0.144 &   0.386  \\ 
  0.00 &  -0.033 &  -0.011 &   4.463 &   5.599 &   2.964 &   1.644 &   4.380 &   3.808 &   1.186 &   6.458 &   0.148 &   0.399  \\ 
  0.10 &  -0.026 &  -0.005 &   4.561 &   5.749 &   3.237 &   1.732 &   4.534 &   4.167 &   1.155 &   6.751 &   0.151 &   0.411  \\ 
  0.20 &  -0.016 &   0.003 &   4.645 &   5.880 &   3.567 &   1.835 &   4.696 &   4.536 &   1.121 &   7.067 &   0.152 &   0.421  \\ 
  0.30 &  -0.003 &   0.015 &   4.746 &   5.974 &   3.971 &   1.961 &   4.874 &   4.888 &   1.079 &   7.392 &   0.156 &   0.434  \\ 
\hline
 [Fe/H] & Mg\,$b$ & Fe5270 & Fe5335 & Fe5406 & Fe5709 & Fe5782 & NaD & H$\gamma_A$ & H$\gamma_F$ & H$\delta_A$ & H$\delta_F$ & \\
\hline
 -2.50 &   0.689 &   0.031 &   0.715 &   0.261 &   0.097 &  -0.214 &   1.054 &   3.152 &   3.079 &   4.586 &   3.776 &\\
 -2.40 &   0.767 &   0.139 &   0.697 &   0.250 &   0.120 &  -0.175 &   1.078 &   3.138 &   3.103 &   4.658 &   3.783 &\\
 -2.30 &   0.855 &   0.239 &   0.673 &   0.236 &   0.142 &  -0.140 &   1.097 &   3.142 &   3.098 &   4.759 &   3.772 &\\
 -2.20 &   0.952 &   0.339 &   0.655 &   0.228 &   0.164 &  -0.104 &   1.127 &   3.146 &   3.086 &   4.872 &   3.764 &\\
 -2.10 &   1.057 &   0.444 &   0.650 &   0.231 &   0.187 &  -0.067 &   1.176 &   3.139 &   3.081 &   4.974 &   3.756 &\\
 -2.00 &   1.174 &   0.559 &   0.663 &   0.246 &   0.213 &  -0.027 &   1.250 &   3.050 &   3.048 &   5.011 &   3.722 &\\
 -1.90 &   1.313 &   0.687 &   0.694 &   0.275 &   0.241 &   0.016 &   1.350 &   2.853 &   2.973 &   4.950 &   3.652 &\\
 -1.80 &   1.471 &   0.821 &   0.739 &   0.314 &   0.271 &   0.060 &   1.471 &   2.600 &   2.883 &   4.834 &   3.567 &\\
 -1.70 &   1.643 &   0.952 &   0.792 &   0.357 &   0.300 &   0.104 &   1.607 &   2.381 &   2.827 &   4.732 &   3.498 &\\
 -1.60 &   1.838 &   1.095 &   0.855 &   0.406 &   0.330 &   0.149 &   1.755 &   1.956 &   2.653 &   4.444 &   3.306 &\\
 -1.50 &   2.059 &   1.243 &   0.925 &   0.458 &   0.361 &   0.192 &   1.909 &   1.296 &   2.318 &   3.947 &   2.968 &\\
 -1.40 &   2.324 &   1.399 &   1.004 &   0.516 &   0.395 &   0.235 &   2.072 &   0.501 &   1.891 &   3.361 &   2.559 &\\
 -1.30 &   2.653 &   1.562 &   1.091 &   0.581 &   0.433 &   0.280 &   2.260 &  -0.373 &   1.428 &   2.795 &   2.166 &\\
 -1.20 &   3.022 &   1.725 &   1.175 &   0.649 &   0.474 &   0.324 &   2.476 &  -1.120 &   1.041 &   2.411 &   1.910 &\\
 -1.10 &   3.394 &   1.857 &   1.245 &   0.709 &   0.515 &   0.363 &   2.701 &  -1.599 &   0.785 &   2.199 &   1.784 &\\
 -1.00 &   3.751 &   1.957 &   1.296 &   0.757 &   0.553 &   0.397 &   2.925 &  -1.834 &   0.641 &   2.155 &   1.756 &\\
 -0.90 &   4.110 &   2.033 &   1.322 &   0.795 &   0.594 &   0.424 &   3.117 &  -2.035 &   0.508 &   2.199 &   1.763 &\\
 -0.80 &   4.468 &   2.102 &   1.340 &   0.836 &   0.638 &   0.455 &   3.278 &  -2.223 &   0.384 &   2.280 &   1.794 &\\
 -0.70 &   4.802 &   2.194 &   1.405 &   0.899 &   0.680 &   0.499 &   3.528 &  -2.273 &   0.308 &   2.318 &   1.822 &\\
 -0.60 &   5.116 &   2.284 &   1.476 &   0.962 &   0.729 &   0.542 &   3.790 &  -2.324 &   0.231 &   2.357 &   1.842 &\\
 -0.50 &   5.416 &   2.377 &   1.556 &   1.029 &   0.782 &   0.586 &   4.061 &  -2.400 &   0.136 &   2.376 &   1.846 &\\
 -0.40 &   5.712 &   2.468 &   1.635 &   1.094 &   0.838 &   0.629 &   4.362 &  -2.445 &   0.041 &   2.443 &   1.858 &\\
 -0.30 &   5.978 &   2.568 &   1.722 &   1.165 &   0.901 &   0.674 &   4.680 &  -2.539 &  -0.072 &   2.441 &   1.853 &\\
 -0.20 &   6.224 &   2.673 &   1.816 &   1.237 &   0.966 &   0.719 &   5.031 &  -2.645 &  -0.193 &   2.396 &   1.841 &\\
 -0.10 &   6.460 &   2.788 &   1.925 &   1.315 &   1.031 &   0.765 &   5.392 &  -2.789 &  -0.319 &   2.289 &   1.816 &\\
  0.00 &   6.677 &   2.913 &   2.049 &   1.400 &   1.098 &   0.812 &   5.760 &  -2.995 &  -0.462 &   2.088 &   1.767 &\\
  0.10 &   6.894 &   3.041 &   2.198 &   1.494 &   1.154 &   0.857 &   6.127 &  -3.209 &  -0.605 &   1.843 &   1.701 &\\
  0.20 &   7.111 &   3.190 &   2.378 &   1.610 &   1.217 &   0.911 &   6.505 &  -3.488 &  -0.767 &   1.523 &   1.617 &\\
  0.30 &   7.330 &   3.364 &   2.599 &   1.756 &   1.292 &   0.980 &   6.918 &  -3.864 &  -0.960 &   1.088 &   1.499 &\\
\enddata
\tablecomments{The entire data of Table~\ref{tab.1.7} are available at http://web.yonsei.ac.kr/cosmic/data/YEPS.htm.}
\end{deluxetable}

\end{document}